\documentclass[prl,nofootinbib,twocolumn,superscriptaddress,preprintnumbers]{revtex4-1}
\usepackage{placeins}
\usepackage{soul}

\let\counterwithin\relax
\usepackage{chngcntr}
\usepackage{color}
\usepackage{epsfig}  
\usepackage{graphicx}
\usepackage{tabularx}
\usepackage{xspace}
\usepackage{float}

\usepackage{color}
\usepackage{hyperref}
\hypersetup{
     colorlinks   = true,
     citecolor    = red,
	 linkcolor=blue
}
\usepackage{halloweenmath}
\usepackage{verbatim}
\usepackage{amsmath}
\usepackage{amssymb}
\usepackage{url}
\usepackage{bbold}
\usepackage{slashed}
\usepackage{array}

\usepackage{multirow}
\usepackage{threeparttable}
\usepackage{paralist}
\usepackage{xspace}
\usepackage{upgreek}
\usepackage{lipsum}

\newcommand{\abs}[1]{\left| #1 \right|} 

\definecolor{darkmagenta}{rgb}{0.55, 0.0, 0.55}
\newcommand{\beq}{\begin{equation}}
\newcommand{\beqn}{\begin{eqnarray}}
\newcommand{\eeq}{\end{equation}}
\newcommand{\eeqn}{\end{eqnarray}}
 
\newcommand\numberthis{\addtocounter{equation}{1}\tag{\theequation}}
\newcommand{\dbar}{\ensuremath{\mathchar'26\mkern-12mu d}}

\newcommand{\jcap}{JCAP}

\DeclareMathAlphabet\mathbfcal{OMS}{cmsy}{b}{n}
\let\arrowedvec=\vec 
\renewcommand{\vec}[1]{\mathbf{#1}}

\newcommand{\msol}{\, \text{M}_\odot}

\newcommand{\ourtitle}{The cosmology of sub-MeV dark matter freeze-in}

\definecolor{blue}{rgb}{0.1,0.4,0.6}

\def\mysection#1{{\it #1.---} }
\begin{document}
\preprint{MIT-CTP/5256}
\title{\ourtitle \vspace*{-0.3cm}}
\author{Cora Dvorkin}
\affiliation{Department of Physics, Harvard University, Cambridge, MA 02138, USA}
\email{cdvorkin@g.harvard.edu}
\author{Tongyan Lin}
\affiliation{Department of Physics, University of California, San Diego, CA 92093, USA}
\email{tongyan@physics.ucsd.edu}
\author{Katelin Schutz}
\affiliation{Center for Theoretical Physics, Massachusetts Institute of Technology, Cambridge, MA 02139, USA}
\email{kschutz@mit.edu}

\begin{abstract} \noindent 
Dark matter (DM) could be a relic of freeze-in through a light mediator, where the DM is produced by extremely feeble, IR-dominated processes in the thermal Standard Model plasma. In the simplest viable models with the DM mass below the MeV scale, the DM has a small effective electric charge and is born with a nonthermal phase-space distribution. This DM candidate would cause observable departures from standard cosmological evolution. In this work, we combine data from the cosmic microwave background (CMB), Lyman-$\alpha$ forest, quasar lensing, stellar streams, and Milky Way satellite abundances to set a lower limit on freeze-in DM masses up to $\sim 20\,$keV, with the exact constraint depending on whether the DM thermalizes in its own sector. We perform forecasts for the CMB-S4 experiment, the Hydrogen Epoch of Reionization Array, and the Vera Rubin Observatory, finding that freeze-in DM masses up to $\sim 80\,$keV can be explored. These cosmological probes are highly complementary with proposed direct-detection efforts to search for this DM candidate. 
\end{abstract}
\maketitle

\mysection{Introduction}
Despite the abundant evidence of dark matter (DM) in our Universe, its fundamental properties and early-universe origins remain open questions. If DM is a particle, it may arise from thermal processes in the primordial plasma in the first moments after the Big Bang. In the scenario known as \emph{freeze-in}, DM is produced from the annihilation or decay of Standard Model (SM) particles in the early universe~\cite{Asaka:2005cn,Asaka:2006fs,Gopalakrishna:2006kr,Page:2007sh,Hall:2009bx,Bernal:2017kxu}. As the Universe cools, the interactions that make DM become inefficient, yielding a fixed DM relic density that is observed at the present day.

If the DM and the force-carrier particle that mediates freeze-in are sufficiently light, then on dimensional grounds the rate for SM particles to produce DM via an $s$-wave process must scale like $\Gamma \sim g_\chi^2 g_{\rm SM}^2 T$ for a relativistic plasma of temperature $T$, where $g_\chi$ is the DM-mediator coupling and $g_{\rm SM}$ is the SM-mediator coupling. Meanwhile, the Hubble rate scales like $H\sim T^2/M_\text{Pl}$ where $M_\text{Pl}$ is the Planck mass. This scaling indicates that freeze-in will predominantly occur at the lowest kinematically accessible temperatures, meaning that in the absence of additional interactions, \emph{the relic DM abundance produced during freeze-in is independent of initial conditions}. Producing the observed DM relic abundance implies a tiny value for the coupling constants, making the parameter space difficult to target with accelerator searches. However, the light mediator greatly enhances the signal of this candidate in direct-detection experiments, since scattering via a light mediator scales like $v^{-4}$ for velocity $v$, which is $v\sim 10^{-3}c$ at the Earth's location in the Milky Way (MW). In the light-mediator regime, the requisite DM-SM couplings for the observed DM relic abundance provide a highly predictive benchmark for recently proposed sub-GeV DM direct-detection experiments~\cite{Essig:2011nj,Essig:2012yx,Essig:2015cda,Hochberg:2016ntt,Derenzo:2016fse,Hochberg:2017wce,Knapen:2017ekk,Griffin:2018bjn,Schutz:2016tid,Knapen:2016cue,Hochberg:2015pha,Hochberg:2015fth,Hochberg:2019cyy,Kurinsky:2019pgb,Griffin:2019mvc,Coskuner:2019odd,Geilhufe:2019ndy,Berlin:2019uco,Griffin:2020lgd}. 

There are strong stellar emission and fifth force constraints on many kinds of light mediators coupled to the SM~\cite{Knapen:2017xzo,Green:2017ybv}. The only light mediators that can be responsible for freeze-in of DM masses below 1~MeV are the SM photon or an ultralight kinetically mixed dark photon. Thus, DM made by freeze-in below 1~MeV will effectively have a small electromagnetic charge, $Q = g_\chi g_{\rm SM}/e \sim 10^{-11}$, defined relative to the electron charge $e$. \emph{Freeze-in is the simplest allowed way to make charged DM}, since the charges required for DM production via freeze-out are excluded by many orders of magnitude~\cite{McDermott:2010pa}. Charged DM has recently been the subject of keen interest in the context of the anomalous observation by the Experiment to Detect the Global EoR Signature (EDGES)~\cite{Bowman:2018yin,Barkana:2018cct,Berlin:2018sjs} and can also play a role in energy loss from stellar and supernova environments \cite{Davidson:2000hf,Vogel:2013raa, Chang:2018rso} and gas clouds~\cite{Bhoonah:2018gjb,Wadekar:2019xnf}, as well as potentially leading to novel plasma behavior in galaxies and clusters~\cite{Ackerman:mha,Heikinheimo:2015kra,Stebbins:2019xjr,Li:2020wyl,Dunsky:2018mqs,Lasenby:2020rlf}. The scenario involving a dark photon is also of theoretical interest, as ultralight bosons are generically expected as states in the spectrum of various string theories~\cite{Abel:2008ai,Goodsell:2009xc}. 

Because of the extraordinarily small couplings involved, freeze-in DM never achieves a thermal number density in the early universe. This means that \emph{freeze-in is one of the few allowed ways of making sub-MeV DM from the SM thermal bath}. Most other mechanisms to produce sub-MeV DM from the thermal bath are excluded (there are some exceptions to this, see e.g. \cite{davidson2000updated,Shi:1998km,Berlin:2017ftj}) because the sub-MeV DM would carry substantial energy and entropy density which would observably alter the effective number of relativistic degrees of freedom, $N_\text{eff}$, and Big Bang Nucleosynthesis (BBN) (see e.g. Refs.~\cite{Boehm:2013jpa,Nollett:2013pwa,Green:2017ybv,Knapen:2017xzo,Sabti:2019mhn}). Note that ultralight dark photon mediators are not produced abundantly by the SM bath in the early universe because of an in-medium suppression of the coupling~\cite{An:2013yfc}, which means the dark photons do not affect $N_\text{eff}$ or BBN.

Sub-MeV freeze-in via an ultralight vector mediator poses a well-motivated DM theory with a complete and consistent early-universe thermal history and a host of concrete predictions for observable phenomena. In this \emph{Letter}, we explore the effects of this production mechanism on the subsequent cosmology, focusing on two key effects: (1) the portal responsible for creating the DM necessarily implies that there is a drag between the DM and the photon-baryon fluid before and during recombination, altering the anisotropies seen in the cosmic microwave background (CMB) and (2) the DM is born with a nonthermal, high-velocity phase-space distribution, which prevents it from clustering on small scales. We constrain the former effect with the \emph{Planck}~2018 CMB power spectra and show how the bound can improve with the planned CMB-S4 experiment. We consider the latter effect in the context of the Lyman-$\alpha$ forest, strong gravitational lensing of quasars, stellar streams, and MW satellites analyzed by the Dark Energy Survey (DES) collaboration. We additionally forecast the DM masses that can be explored with observations of the 21~cm power spectrum with the Hydrogen Epoch of Reionization Array (HERA), and of the subhalo mass function with the Vera Rubin Observatory. 

For both observable effects, the full velocity distribution of the DM is of critical relevance. The DM-SM scattering cross section responsible for the drag effect scales like $v^{-4}$ and depends strongly on the low-velocity part of the distribution, while the suppressed growth of structure is sensitive to DM in the high-velocity tail of the distribution. In this work, we use the phase space derived in Ref.~\cite{Dvorkin:2019zdi} which is highly nonthermal at production, although the distribution could become thermal prior to recombination through DM-DM interactions. Here we consider both the nonthermal and thermalized phase space, which bookend the range of intermediate possibilities. Our results are summarized in Fig.~\ref{fig:constraints}.

\begin{figure}[t]
 \includegraphics[width = 0.49\textwidth]{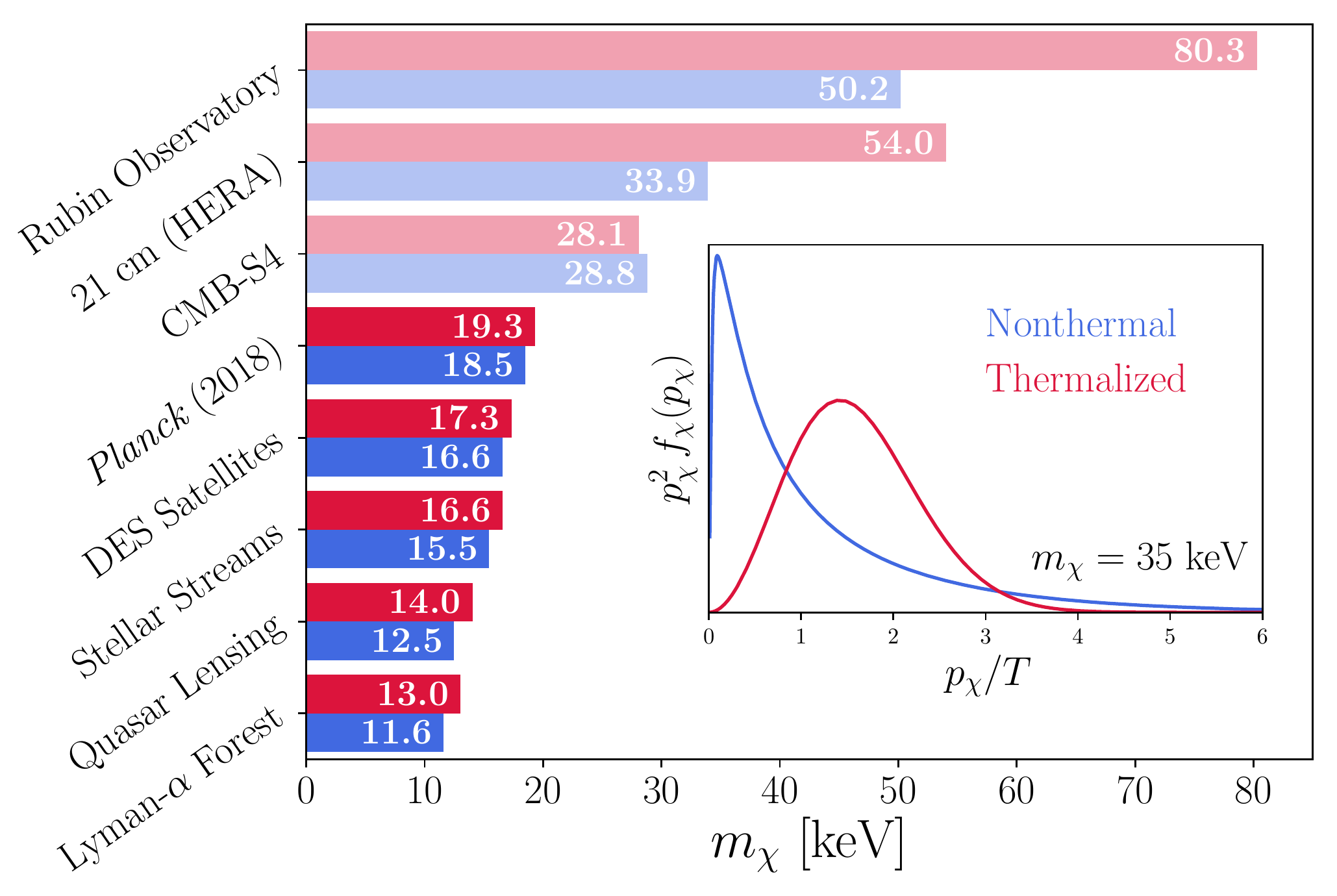}
 \vspace{-0.7cm}
\caption{Cosmological 95\% bounds on DM that is produced by freeze-in through a vector mediator. Dark shaded bars correspond to excluded DM masses while light ones correspond to projected future reach. These differ depending on whether the DM phase space remains nonthermal (blue) or thermalizes through self-scattering (red). Inset: the DM phase space.} 
\label{fig:constraints}
\vspace{-0.6cm}
\end{figure}

\mysection{Particle Properties}
Throughout this work we assume that DM is a Dirac fermion and we focus on the keV-MeV range for the DM mass, $m_\chi$. We assume that the DM couples to the SM photon, either (1) at the level of the Lagrangian with coupling strength $eQ$ or (2) effectively through an ultralight (sub-eV) dark vector portal, $A'$, where $eQ$ is a product of the dark $U(1)'$ gauge coupling $g_\chi$ and the kinetic mixing parameter $\kappa$. 

In the keV-MeV mass range, two channels are dominantly responsible for the production of DM: electron-positron annihilation $e^+ e^- \to \chi \bar \chi$ with thermally averaged cross section $\langle\sigma v\rangle_\text{ann.}$ and plasmon decays $\gamma^* \to \chi \bar \chi$ with rate $\langle \Gamma\rangle_\text{plas.}$ (which includes a thermal boost factor). In the absence of additional interactions, a cosmologically relevant amount of DM will \emph{not} be produced in the very early universe for small SM-DM couplings due to the sub-Hubble rates for these processes at high $T$. We therefore assume a negligible initial abundance of DM. For freeze-in production of DM, it is possible to semi-analytically solve the Boltzmann equation, $\partial{f}_\chi/\partial t - H (p_\chi^2/E_\chi )\partial f_\chi /\partial E_\chi = C[f_e, f_{\gamma^*}]/E_\chi$, for the DM phase space $f_\chi$. The collision term $C$ does not depend on $f_\chi$ to a very good approximation due to the small DM number density relative to the number density of particles in the SM plasma (in other words, one can safely ignore any DM backreactions). In Ref.~\cite{Dvorkin:2019zdi}, we integrated this equation to find the DM phase-space distribution from freeze-in. The typical freeze-in DM momentum is of order the photon temperature (see Fig.~\ref{fig:constraints}) since the DM inherits the kinematic properties of the plasma from which it is born.

\begin{figure*}[ht!]
\includegraphics[width=0.98\textwidth]{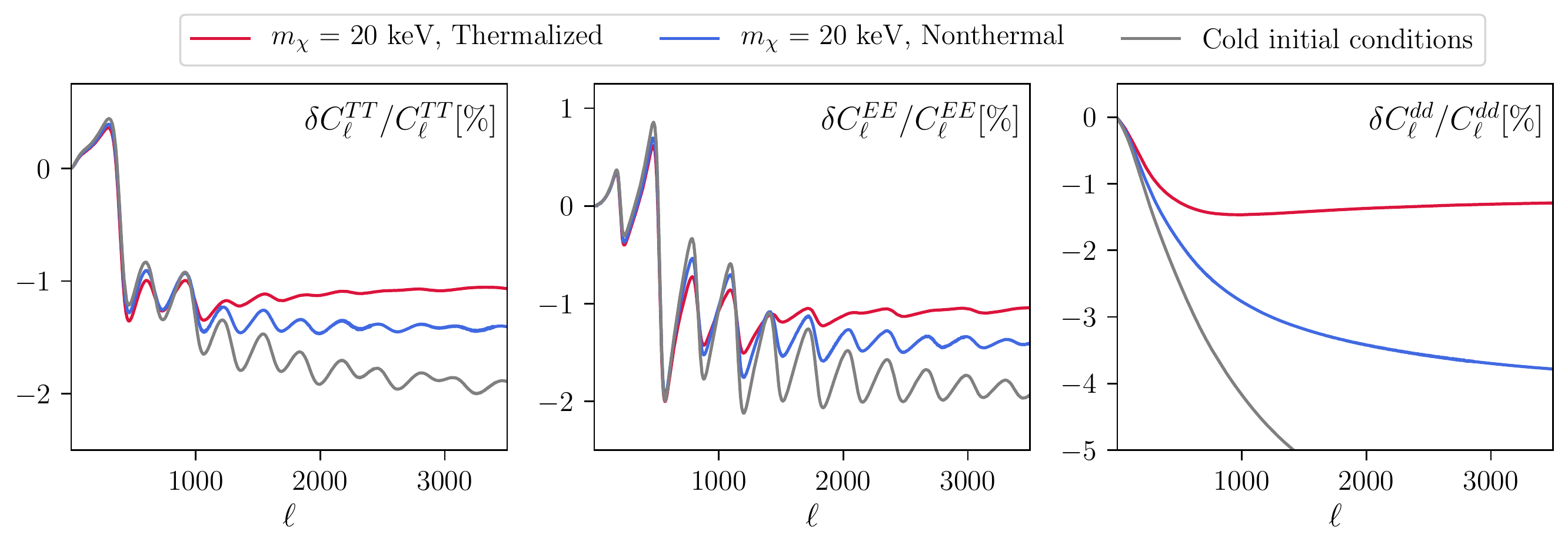}
\vspace{-0.6cm}
\caption{Effect of DM-baryon drag on CMB temperature, polarization, and lensing power spectra. We show freeze-in DM with mass of 20 keV for different DM thermal histories. We also show the effect for DM that has the same cross section and cold initial conditions, which is most similar to previous studies in Refs.~\cite{Dvorkin:2013cea,Boddy:2018wzy,Slatyer:2018aqg}.}
\vspace{-0.5cm}
\label{fig:delta_Cls}
\end{figure*}

The relic DM abundance is determined by the 0th moment of the Boltzmann equation, given by $d n_\chi /d t + 3 H n_\chi = n_e^2 \langle\sigma v\rangle_\text{ann.} + n_{\gamma^*} \langle \Gamma\rangle_\text{plas.}$, where $n_i$ is the number density of species $i$. In this work, we assume that all DM is produced by freeze-in. Solving this equation uniquely determines the effective charge $Q$ for a given DM mass, where $Q \sim 10^{-11}$ for the range of masses considered here. For $Q$ and $m_\chi$ saturating the relic abundance, we can also determine the entire DM phase-space distribution at the time of production and subsequent effects on cosmology. Because the cosmological effects depend on DM phase space, the limits described in this \emph{Letter} do not apply for arbitrary $Q$ and $m_\chi$. For instance, if $Q\gtrsim 10^{-10}$ for DM in the keV-MeV mass range, DM would be overproduced by freeze-in. Additional features of the model would be required to subsequently deplete the DM abundance, which would necessarily impact the phase space and cosmological observables.

While integrating the Boltzmann equation gives us the phase-space distribution when DM is produced, the effects we study are sensitive to the DM velocity at later times, and DM-DM self-scattering can redistribute the phase space. In our analysis, we consider two limiting cases for the DM phase-space distribution: (1) the fully nonthermal primordial phase space and (2) a Gaussian phase space for DM that has thermalized within its own isolated sector through DM-DM interactions, which leads to a temperature $T_\chi$ that preserves $\left<p_\chi^2\right>$ (assuming thermalization occurs while the DM is nonrelativistic). The first case occurs if the mediator is the SM photon, in which case the self-scattering via the SM photon is highly suppressed as $Q^4$. If the mediator is a dark photon, DM-DM scattering is only suppressed as $g_\chi^4$ and thermalization of the DM phase space can occur more efficiently. DM self-thermalization requires a high value of $g_\chi$, which can be compensated by lowering $\kappa$ to give the same value of $Q$. Due to bounds on DM self-interaction (for instance from merging galaxy clusters, see e.g.~\cite{Tulin:2017ara}), $g_\chi$ cannot be too large. However, DM can potentially self-thermalize as early as redshift $z\sim 10^6$ without violating self-interaction bounds~\cite{Dvorkin:2019zdi}. Because thermalization is not instantaneous, a given value of $g_\chi$ implies some time-dependent DM phase space. The bounds we present are meant to serve as endpoints of the parameter space, while in the intermediate regime there may be other effective descriptions of the phase space~\cite{Huo:2019bjf}.

\mysection{Baryon Dragging}
The portal responsible for making DM gives rise to a DM-baryon scattering cross section scaling as $v^{-4}$, which results in a non-gravitational drag force between the DM and the photon-baryon fluids. The drag force introduces extra damping in the amplitude of acoustic oscillations and causes a slight suppression in the matter power spectrum for modes that are inside the horizon while the drag is most active. We note that this does not add significant constraining power compared to the effect of the high-velocity DM phase space, which we explore in the next section and Supplemental Materials.

We calculate the effects of DM-baryon drag on the CMB, shown in Fig.~\ref{fig:delta_Cls}, using a modified version of the Boltzmann solver \texttt{CAMB} \cite{Lewis:1999bs,2012JCAP...04..027H} with additional terms in the Boltzmann equations. Further details are given in the Supplemental Material. In this work, we assume DM only scatters with protons and neglect DM-helium and DM-electron scattering; DM-helium drag is suppressed relative to DM-proton drag because the $v^{-4}$ scattering is cut short by earlier helium recombination, while DM-electron drag is suppressed due to the low mass and high thermal velocity of electrons. 
The effect of the drag is that the first acoustic peak in the CMB is enhanced while higher-$\ell$ fluctuations in power spectra are suppressed. The constraints are driven by the suppression at higher $\ell$. For freeze-in, the high-velocity DM phase space leads to a smaller drag rate and correspondingly smaller $\delta C_\ell$ compared to DM-baryon scattering with cold initial conditions. Among the freeze-in thermal histories, the nonthermal case has more low-velocity DM particles, resulting in larger drag rate and $\delta C_\ell$ than the thermalized case.

We set constraints by running a Markov Chain Monte Carlo likelihood analysis, using CMB temperature, polarization, and lensing data from the \emph{Planck}~2018 release~\cite{Aghanim:2019ame}. For fixed DM mass, we vary the six standard $\Lambda$CDM parameters in addition to the normalization of the DM-baryon drag. The lower bound on the freeze-in mass is then found by interpolating the constraints on DM-baryon drag for a few masses, and finding where the normalization matches that of freeze-in. The primary degeneracy with $\Lambda$CDM is with the scalar spectral index $n_s$, since changing $n_s$ can also result a suppression of the acoustic peaks at high $\ell$. This degeneracy is slightly larger for the nonthermal phase space, which is why we find a weaker {\it Planck} constraint despite the larger $\delta C_\ell$. 

To project the sensitivity of the future CMB-S4 experiment~\cite{Abazajian:2016yjj}, we perform Fisher forecasts with the unlensed CMB $TT, EE,$ and $TE$ spectra in addition to the lensing deflection spectrum $dd$. Assuming that CMB-S4 can be combined with {\emph{Planck}} data, we take a minimum multipole $\ell_{\rm min} = 30$ and impose a prior on the optical depth $\tau = 0.06 \pm 0.01$. We take a fractional sky coverage of $f_{\rm sky} = 0.4$, a maximum multipole of $\ell_{\rm max} = 5000$ for temperature and polarization (except for $TT$ where $\ell_{\rm max} = 3000$) and $\ell_{\rm max} = 2500$ for lensing. We consider noise levels corresponding to a beam resolution of $\theta_{\rm FWHM} = 1$ arcmin and a noise temperature of 1 $\mu$K-arcmin in temperature and $\sqrt{2}$ $\mu$K-arcmin in polarization. For the lensing power spectrum $C_\ell^{dd}$, the noise curves are obtained from a procedure of iterative delensing using E-modes and B-modes \cite{Abazajian:2016yjj}. CMB lensing reduces the degeneracy with $n_s$ and provides additional constraining power on the nonthermal case, which is the primary factor that drives the stronger forecast in comparison to the thermal case. This is because lensing is relatively more powerful in constraining DM-baryon scattering at higher redshifts~\cite{Li:2018zdm,Boddy:2018wzy}, which is larger in the nonthermal case. The effects of DM-baryon drag are mildly degenerate in the primary CMB with effects beyond the six $\Lambda$CDM parameters, for instance massive neutrinos and $N_\text{eff}$; however, this degeneracy will be broken by future measurements of CMB lensing.

Beyond the CMB, DM-baryon drag with a scattering cross section $\propto v^{-4}$ has been proposed to explain an anomalous 21~cm absorption trough seen in EDGES~\cite{Bowman:2018yin}, and could be searched for with the 21~cm power spectrum~\cite{Munoz:2015bca,Munoz:2018jwq}. However, for freeze-in, we constrain the cross section to be too small to explain the absorption seen in EDGES~\cite{Barkana:2018cct}, and it would also be challenging to see in the 21~cm power spectrum~\cite{Munoz:2018jwq}.

\mysection{Effect on clustering}
Because the phase space of freeze-in DM is initially inherited from relativistic electron-positron pairs and plasmons, sub-MeV DM is produced with a relatively high-speed phase-space distribution, leading to a suppression in gravitational clustering on small scales, as shown in Fig.~\ref{fig:pspecs}.

\begin{figure}[t]
\includegraphics[width=0.49\textwidth]{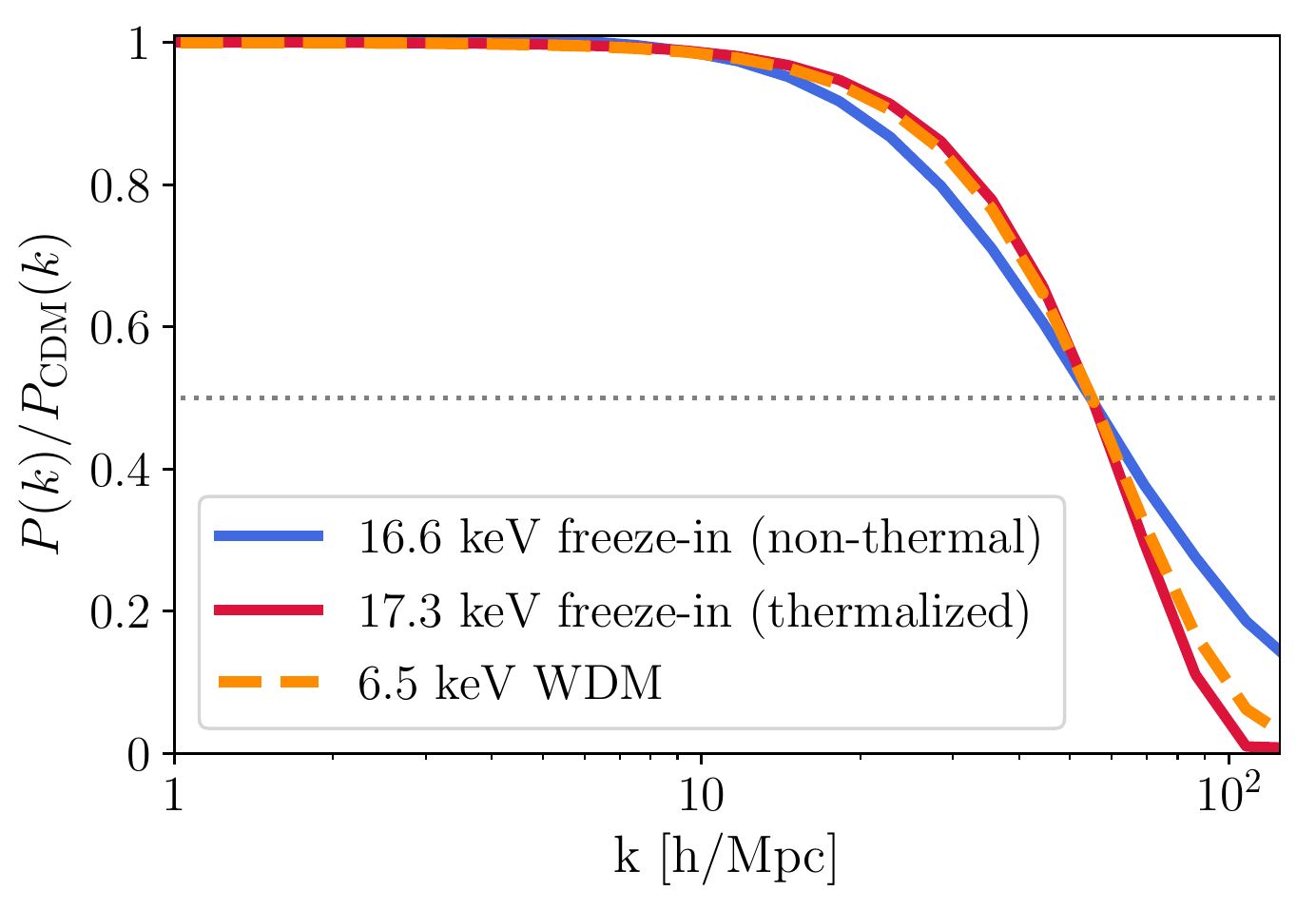}
\vspace{-0.9cm}
\caption{Suppression of the linear matter power spectrum relative to CDM, for different DM thermal histories. 
For the WDM case, we show a mass of 6.5~keV which corresponds to the current strongest limits from DES~\cite{Nadler:2020prv}. For the freeze-in scenarios, we show DM masses which match the WDM power spectrum at the scale where power is suppressed by half.}
\vspace{-0.6cm}
\label{fig:pspecs}
\end{figure}

The clustering properties of freeze-in DM depend on whether the DM retains its nonthermal phase-space distribution. For the case where DM self-thermalizes and obtains a temperature $T_\chi$ at early times, the suppression can be characterized by an effective sound speed for the DM fluid, $c_\chi^2 = 5 T_\chi/3m_\chi$. We then calculate the matter power spectrum using the fluid equations given in the Supplemental Materials, and implemented in \texttt{CAMB}. For the nonthermal phase space, the suppression arises due to free-streaming and cannot be described by fluid equations (i.e. the DM free-streaming length is longer than the mean free path). In this case, we compute the transfer functions for the linear matter power spectrum using the Boltzmann code \texttt{CLASS}~\cite{2011JCAP...07..034B}, treating the DM as a massive neutrino species with phase-space distribution determined by freeze-in.

There are a variety of astrophysical systems that probe clustering on scales where freeze-in would have an effect. To set the limits on freeze-in shown in Fig.~\ref{fig:constraints}, we compare these transfer functions with those of the well-studied case of warm DM (WDM). The WDM power spectrum as computed with Boltzmann solvers is well described by the fitting form~$P_L^\text{WDM}= P_L^\text{CDM} \left[ 1 + (\lambda_\text{fs}^\text{eff} k)^{2 \nu})\right]^{-10/\nu}$
where $P_L$ is the linear matter power spectrum, $\nu = 1.12$, and the effective free-streaming length is $\lambda_\text{fs}^\text{eff}  = 0.07\times (m_\text{WDM} / 1~\text{keV})^{-1.11} \,$Mpc assuming WDM of mass $m_\text{WDM}$ comprises all the DM and a Hubble parameter of $H_0=70$~km/s/Mpc~\cite{bode2001halo,viel2005constraining,Schneider:2011yu}. In Fig.~\ref{fig:pspecs} we compare the suppression in the linear matter power spectrum for freeze-in and WDM. We have fixed the DM masses to have identical half-mode scales, {\emph{i.e.}}, we match the scale $\lambda_\text{1/2}$ where the power spectrum is suppressed by half compared to CDM. For WDM, the half-mode scale is $\lambda_\text{1/2} = 2 \pi \lambda_\text{fs}^\text{eff} (2^{\nu/10} - 1)^{-1/2 \nu} \approx 19.33  \lambda_\text{fs}^\text{eff}$. 

In the standard WDM scenario, DM is a thermal relic that is produced relativistically and the DM temperature is fixed to saturate the observed relic abundance, $T_{\rm WDM} = 0.16 \times(1\,\text{keV}/m_\text{WDM})^{1/3}\, T_\gamma$ (note that this temperature difference requires $\mathcal{O}(10^3)$ degrees of freedom in the early universe, whereas the entire SM has 106.75). DM produced by freeze-in has a higher effective temperature. We therefore find that the half-mode scale for freeze-in at a given mass matches that of WDM with a smaller mass. Freeze-in DM that is able to self-thermalize yields a transfer function that almost exactly matches WDM for the appropriate choice of masses. However, there is a difference in the shape of the transfer function for the nonthermal freeze-in phase space, which has a larger high-velocity tail. We consider limits on the nonthermal freeze-in phase space to be estimates, and expect that more accurate limits can be obtained by accounting for the full transfer function in simulations.

One of the most well-studied tracers of the the matter power spectrum on scales $k\sim 10 \,h/$Mpc is the Lyman-$\alpha$ forest of quasar spectral absorption lines. We use the quoted 95\% C.L. limit of $m_{\rm WDM} > 5.3$ keV from Ref.~\cite{Irsic:2017ixq} which comes from a combined analysis of data from the XQ-100 survey observed with the X-shooter spectrograph~\cite{lopez2016xq} and from the HIRES/MIKE spectrographs~\cite{Viel:2013apy} in tandem with hydrodynamical simulations of the intergalactic medium. Refs.~\cite{Yeche:2017upn,Baur:2017stq} found a similar constraint from the combined analysis of XQ-100, HIRES/MIKE data below $z=5$, and data from the Baryon Oscillation Spectroscopic Survey of the Sloan Digital Sky Survey (SDSS)~\cite{Palanque-Delabrouille:2013gaa}. For the case where DM cannot self-thermalize, we follow Ref.~\cite{Murgia:2018now} which examined a variety of shapes in the suppression of the matter power spectrum (corresponding to alternative theories of DM beyond WDM) in the context of limits from Lyman-$\alpha$ flux power spectra. A general trend found in this work is that for fixed half-mode scale, power spectra with a shallower decline are more readily rejected by analyses of the Lyman-$\alpha$ forest than steeper counterparts. It is therefore conservative to apply a limit on nonthermal freeze-in phase space by matching the half-mode scale. Based on improved recent constraints on the ultralight DM transfer function~\cite{Hu:2000ke,Hlozek:2016lzm} from the Lyman-$\alpha$ forest~\cite{2020arXiv200712705R}, we estimate that a constraint on freeze-in of $m_\chi \sim 30\,$keV can be set in the near future with similar methods. 

Additional limits come from the abundance of DM halos and subhalos, which inherits any small-scale suppression in the matter power spectrum. In particular, the half-mode scale $\lambda_{1/2}$ translates to a halo mass scale $M_{1/2} = \pi \lambda_{1/2}^3 \bar{\rho}_m/6$, corresponding to the average mass contained within a spatial region of size $\lambda_{1/2}$. The halo and subhalo mass functions (i.e., abundances as a function of mass) are suppressed for masses much smaller than $M_{1/2}$, meaning that the discovery of low-mass subhalos excludes WDM below some particle mass scale. Ref.~\cite{Nadler:2019zrb} considered the population of classical and SDSS-discovered MW satellites, finding a constraint of $m_{\rm WDM} > 3.3$~keV. This constraint was strengthened to $m_{\rm WDM} > 6.5$~keV with the inclusion of MW satellites from DES and Pan-STARRS (PS)~\cite{Nadler:2020prv}. Studies of strong gravitational lensing of quadruply imaged quasars by foreground galaxies provide further evidence for an abundance of subhalos. The presence of subhalos alters the flux ratios and positions of the lensed images in a way that depends on combinations of the second derivative of the lensing potential; analyses of such systems constrain $m_{\rm WDM} > 5.6$~keV~\cite{hsueh2020sharp} (see also~\cite{Gilman:2019nap}). The combination of satellite abundances and quasar lensing with the Lyman-$\alpha$ forest can improve the constraint slightly, yielding a combined limit of 6.7~keV~\cite{Enzi:2020ieg}. The presence of subhalos can also perturb the densities of stellar streams in a characteristic way, leaving gaps in the stream that can persist on gigayear timescales. Refs.~\cite{Banik:2019cza,Banik:2019smi} claim evidence for DM substructure based on an analysis of the GD-1 and Pal~5 streams as observed by PS and \emph{Gaia}; when combined with classical MW satellite abundances, perturbations in stellar streams constrain $m_\text{WDM}> 6.3$~keV.

In the future, lower mass halos can be probed via the 21~cm absorption signal from cosmic dawn; low-mass halos are the first to form stars, whose radiation initially couples the spin temperature of neutral hydrogen to the kinetic gas temperature through the Wouothuysen-Field effect and later heats the intergalactic medium. By considering how small-scale structure formation would affect the star formation history and leave an imprint on the 21~cm power spectrum as seen by HERA, Ref.~\cite{Munoz:2019hjh} found that $m_\text{WDM}\sim14$~keV could be constrained in the near future. Finally, a target of the Rubin Observatory is to probe $m_\text{WDM}\sim 18$~keV by probing the subhalo mass function down to masses of $10^6 \msol$~\cite{Drlica-Wagner:2019xan}.

\mysection{Conclusions} 
Sub-MeV freeze-in via a light vector mediator sits at the nexus of many interesting possible DM properties. Freeze-in is the only minimal way to make charged DM and is one of very few ways to make sub-MeV DM from a SM thermal process in the early universe. These properties, combined with a predictive direct-detection signal in the light-mediator regime, make freeze-in a key benchmark for proposed sub-MeV direct-detection experiments. This DM candidate is dominantly born from the decay of plasmons and has a nonthermal, high-velocity phase-space distribution, making it behave like WDM. Based on observations of clustering on small scales, we have excluded freeze-in masses below $\sim17\,$keV. This DM candidate can also scatter with baryons in the primordial plasma, altering the CMB and allowing us to exclude freeze-in masses below $\sim19\,$keV. In the near future, cosmological probes have substantial room for improvement and will test freeze-in masses up to almost 100~keV, greatly complementing terrestrial efforts to directly detect sub-MeV DM.

\mysection{Acknowledgements}
We thank Asher Berlin, Steen Hansen, Adrian Liu, Pat McDonald, Julian Mu\~noz, Keir Rogers, Tracy R. Slatyer, and Linda Xu for useful conversations pertaining to this work. We acknowledge the importance of equity and inclusion in this work and are committed to advancing such principles in our scientific communities. CD was partially supported by NSF grant AST-1813694. TL is supported by an Alfred P. Sloan foundation fellowship and the Department of Energy under grant DE-SC0019195. KS is supported by a Pappalardo Fellowship and received support from the National Science Foundation and the Hertz Foundation during early stages of this work. 

\bibliography{main}
\clearpage
\newpage
\maketitle
\onecolumngrid

\clearpage
\newpage
\maketitle
\onecolumngrid
\begin{center}
\textbf{\large \ourtitle} \\
\vspace{0.05in}
{ \it \large Supplemental Material}\\
\vspace{0.05in}
{ Cora Dvorkin, Tongyan Lin, Katelin Schutz}
\end{center}
\counterwithin{figure}{section}
\counterwithin{table}{section}
\counterwithin{equation}{section}
\setcounter{equation}{0}
\setcounter{figure}{0}
\setcounter{table}{0}
\setcounter{section}{0}
\renewcommand{\theequation}{S\arabic{equation}}
\renewcommand{\thefigure}{S\arabic{figure}}
\renewcommand{\thetable}{S\arabic{table}}
\newcommand\ptwiddle[1]{\mathord{\mathop{#1}\limits^{\scriptscriptstyle(\sim)}}}

\section{Drag from DM-baryon scattering}

\subsection{Boltzmann equations}
The presence of interactions between DM and SM particles implies that the DM fluid can be dragged non-gravitationally by the photon-baryon plasma at a small level. Here we consider DM-baryon scattering to be the dominant interaction channel, since DM-photon scattering is suppressed by an additional factor of $Q^2$ relative to DM-baryon scattering and does not receive the same low velocity $v^{-4}$ enhancement relevant to DM-baryon scattering. In what follows, we further assume that the dominant mode of DM-baryon scattering is with free protons. Electrons have a much higher thermal velocity due to their low mass, making the $v^{-4}$ enhancement significantly weaker for scattering on electrons (we explicitly find that the drag rate between DM and electrons is suppressed by four orders of magnitude compared to the drag rate between DM and protons). Furthermore, scattering with free helium nucleii is also a subdominant effect ($\sim$10\% level) because of the higher redshift of helium recombination, indicating that DM-helium scattering stops when typical velocities are still relatively high and when the scattering is less enhanced. Neglecting other modes of scattering besides DM-proton scattering is a conservative approximation in the sense that it slightly underpredicts the true degree of scattering and therfore will lead to a slightly weaker constraints from our analysis. In the rest of the Supplemental Material, we will sometimes refer to properties of the baryons collectively as part of one unified plasma; at other times, we will more explicitly refer to properties of protons, for instance inserting relevant factors of the proton mass $m_p$ or factors of the free proton mass fraction, which we denote $X_p$, to keep track of what fraction of particles can scatter and transfer energy and momentum.

To understand the effects of the drag arising from DM-baryon scattering, we consider the bulk acceleration and energy transfer imparted to the DM fluid averaged over many collisions. Working in synchronous gauge~\cite{ma1995cosmological}, and allowing for a nonzero peculiar velocity for DM that comes from the interaction with baryons \cite{Chen:2002yh}, the momentum and energy transfer lead to modified evolution equations for the fluid perturbations: 
\begin{align}
{\dot \delta_\chi} &= -\theta_\chi - \frac{\dot{\mathfrak{h}}}{2}, \\ 
{\dot \delta_b} &= -\theta_b- \frac{\dot{\mathfrak{h}}}{2},  \\
{\dot \theta_\chi} &= -\frac{\dot a}{a}\theta_\chi  + c^2_\chi k^2 \delta_\chi + R_\chi (\theta_b - \theta_\chi ), \\
{\dot \theta_b} &= -\frac{\dot a}{a}\theta_b + c^2_b k^2 \delta_b + R_\gamma (\theta_\gamma - \theta_b)  + \frac{\rho_\chi}{\rho_b} R_\chi (\theta_\chi- \theta_b), \label{baryonspeedchange}
\end{align} 
where overdots are derivatives with respect to conformal time, subscripts denote the species (DM denoted as $\chi$ and baryons as $b$), $\delta$ denotes a perturbation to the mass density $\delta \equiv \rho/\bar{\rho} -1$ of the relevant species, $\theta$ is the divergence of the fluid peculiar motion (relative to Hubble flow) for a given species, and $\mathfrak{h}$ is the trace of the spatial part of the metric perturbation in synchronous gauge. Additionally, $c$ denotes the sound speed of the relevant species, where $c_\chi^2 = 5 T_\chi^{\rm eff}/ 3m_\chi$ for nonrelativistic DM that is decoupled from the baryons and we use the effective temperature for freeze-in. The DM-baryon velocity exchange rate is parameterized as $R_\chi$ (i.e. corresponding to the coupling between the fluids, defined analogously to the usual $R_\gamma=(4\bar{\rho}_\gamma/3 \bar{\rho}_b) a n_e \sigma_{\rm t}$ that couples the photons to the baryons through Thomson scattering with cross section $\sigma_\text{t}$). Note that we define $R_\chi$ to be the rate of change of the bulk DM velocity; to get the change in the baryon velocity, the factor of $\rho_\chi/\rho_b$ is needed in the final term of Eq.~\eqref{baryonspeedchange} to account for the differences in inertia of the fluids. {While the evolution equations above are general, at early times the baryons and photons are tightly coupled; in that epoch, we use the tight coupling approximation with DM-baryon drag derived in Ref.~\cite{Sigurdson:2004zp,Xu:2018efh}.}

We work in the nonrelativistic limit for the DM, since DM-baryon scattering is most active at low velocities due to the light mediator. Then the primary effect of nontrivial DM microphysics is through the velocity exchange rate encapsulated by $R_\chi$, which can be expressed as \beq \frac{d \vec{V}_\text{rel}}{d t} = -  \vec{V}_\text{rel} \frac{R_\chi}{a} \label{Rchidef}\eeq for bulk DM velocity $\vec{V}_\text{rel}$ relative to the baryon fluid (distinct from the velocity of individual particles), where the factor of $a$ accounts for the transformation from conformal time to cosmic time $t$. This bulk DM acceleration can be calculated by computing the 
first moment of the full Boltzmann equation including the collision term. For the DM, this moment reads 
\beq \frac{d \vec{V}_\chi}{dt} +  H \vec{V}_\chi = -\frac{1}{\rho_\chi} \int \frac{ \dbar^3 p_{\chi,i}}{2 E_{\chi,i }}\frac{ \dbar^3 p_{b,i}}{2 E_{b,i}}  \frac{ \dbar^3 p_{\chi,f}}{2 E_{\chi,f}}  \frac{ \dbar^3 p_{b,f}}{2 E_{b,f}} \sum_\text{dof} \abs{\mathcal{M}}^2 (2 \pi)^4 \delta^{(4)}\left(p_{\chi,i}+p_{b,i}- p_{\chi,f} - p_{b,f}\right) \vec{p}_{\chi,i} \left(f_{\chi,i} f_{b,i} - f_{\chi,f} f_{b,f}\right), \label{eq:rawcoll}
\eeq
where the term $H \vec{V}_\chi$ accounts for Hubble damping, the subscript labels $i$ and $f$ refer to DM and baryons scattering in the initial or final state, respectively, and where we have neglected Pauli blocking. Note that the phase-space distributions here are six dimensional, $\int \dbar^3 x ~\dbar^3 p f(\vec{x}, \vec{p}) = 1$. Because the initial and final states' phase-space distributions are identical for the relevant $2 \rightarrow 2$ elastic scattering processes, the momenta can be relabelled and the two terms in the collision integral can be collected to read
\beq \frac{d \vec{V}_\chi}{dt} + H \vec{V}_\chi = \frac{1}{\rho_\chi} \int \frac{ \dbar^3 p_{\chi,i}}{2 E_{\chi,i }} \frac{ \dbar^3 p_{b,i}}{2 E_{b,i}} \frac{ \dbar^3 p_{\chi,f}}{2 E_{\chi,f}}  \frac{ \dbar^3 p_{b,f}}{2 E_{b,f}}\sum_\text{dof} \abs{\mathcal{M}}^2 (2 \pi)^4 \delta^{(4)}\left(p_{\chi,i}+p_{b,i}- p_{\chi,f} - p_{b,f}\right)\Delta \vec{p}_{\chi}  f_{\chi}(p_{\chi,i}) f_{b}(p_{b,i}),  \eeq
where $\Delta \vec{p}_\chi=\vec{p}_{\chi,f}- \vec{p}_{\chi, i}$ is the momentum transfer. Here we have dropped the initial and final labels from the phase space and have assigned those labels to the momentum argument. 

We first focus on performing the integrals over the momenta of the final state particles. Since we are working in the nonrelativistic limit of scattering, we note that the parts of the integrand that depend on final state momenta are Galilean invariant. We therefore have the freedom to perform the integrals over the final state momenta in the center-of-mass (CM) frame as long as any dependence on initial state momenta is written in a frame-independent way. We choose a coordinate system where the initial DM particle is moving in the $\hat{z}$ direction (implying that the baryon it scatters on is moving in the $-\hat{z}$ direction). The momentum transfer in this frame is explicitly $\Delta \vec{p}_{\chi}= p_{CM} (\sin \theta_{CM} \cos \phi_{CM},\sin \theta_{CM} \sin \phi_{CM}, \cos \theta_{CM}-1) $, and $p_{CM} =\mu v_{\rm rel}$ with $\mu$ the DM-baryon reduced mass and $v_{\rm rel}$ is the relative velocity. In the CM frame the differential scattering cross section is given by $d \sigma/d \cos \theta_{CM} = \sum_\text{dof} \abs{\mathcal{M}}^2/32 \pi s$, where the invariant Mandelstam variable $s$ is $s = (E_{\chi,_{CM}}+ E_{b, _{CM}})^2 = \Big(\sqrt{ \smash[b]{m_\chi^2 + p_{CM}^2} } + \sqrt{\smash[b]{m_b^2 + p_{CM}^2} }\Big)^2$. The integration over final state momenta proceeds as 
\begin{align} 
&\frac{8 s}{\pi}\int \frac{d \Omega_{CM} d p_{CM} p_{CM} ^2}{2 E_{\chi,_{CM}} \, 2E_{b,_{CM}}} \frac{d \sigma}{d \cos \theta_{CM}} \delta (\sqrt{s} - E_{\chi,_{CM}}-  E_{b,_{CM}})\Delta \vec{p}_{\chi } \nonumber\\=& -4 \sqrt{s} \int d \cos \theta_{CM} d p_{CM} p_{CM}^2 \frac{d \sigma}{d \cos \theta_{CM}} \delta (p_{CM}  -~ p_0)\, (1- \cos \theta_{CM})\,\hat{z} = - 4 \sqrt{s}\, p_0^2 \sigma_T \hat{z}, \label{finalstates}
\end{align}
where $\sigma_T$ is the momentum transfer cross section for DM-baryon scattering 
\beq \sigma_T  = \int d \cos \theta_{CM}\frac{d \sigma}{d \cos \theta_{CM}} (1- \cos \theta_{CM})\eeq
and $p_{ 0} = \sqrt{(m_\chi^2-m_b^2)^2 - 2 s (m_\chi^2+m_b^2) + s^2} / 2\sqrt{s} $ is the pole of the $\delta$-function. In Eq.~\eqref{finalstates}, we chose a convenient coordinate system with the particles' collision axis pointing in the $\hat{z}$ direction; to relate this to a general coordinate system and frame, we note that in any frame the only vector component of the momentum transfer that survives angular averaging is the component along the scattering axis, which has a Galilean-invariant direction $\hat{v}_\text{rel}$. Noting that in the nonrelativistic limit $p_0^2 \sqrt{s} = (m_b+m_\chi) p_{CM}^2 + \mathcal{O}(p_{CM}^4)$, altogether we have 
\beq \frac{d \vec{V}_\chi}{dt} +  H \vec{V}_\chi \approx 
-\frac{\mu}{\rho_\chi}\int  \dbar^3 p_{\chi,i} ~\dbar^3 p_{b,i} f_{\chi}(p_{\chi,i}) f_{b}(p_{b,i})\, \sigma_T v_\text{rel}^2 \hat{v}_\text{rel}.\label{initialstate} \eeq
If we further assume that all phase-space distributions are separable in configuration space and momentum space, i.e. that $f(\vec{x}, \vec{p})= n(\vec{x}) f_v(\vec{p})$ then we reproduce the well-known expression (see for instance Eq.~A10 of Ref.~\cite{Boddy:2018wzy})
\beq \frac{d \vec{V}_\chi}{dt} +  H \vec{V}_\chi \approx 
-\frac{\rho_b X_p}{ (m_\chi+ m_p)}\int  \dbar^3 p_{\chi} ~\dbar^3 p_{b} f_{\chi, v}(p_{\chi}) f_{b, v}(p_{b})\, \sigma_T v_\text{rel}^2 \hat{v}_\text{rel}, \label{masterdrag}\eeq
where we have dropped the initial state label and have introduced a factor of the free proton mass fraction $X_p$, since only ionized particles are efficient at scattering and since we are assuming the DM-baryon drag is primarily driven by scattering on protons. The momentum transfer cross section for DM-baryon scattering in a freeze-in scenario is 
\begin{align}
    \sigma_T = \frac{4 \pi Q^2 \alpha^2}{\mu^2 v_\text{rel}^4} \ln \left(\frac{2 \mu v_\text{rel}} { m_D}\right),
\end{align} 
where here $\mu$ is the DM-proton reduced mass and $m_D$ is Debye mass, which is given by $m_D = 3.7 \times 10^{-6} T_\gamma$ when the ionization fraction is unity. We will approximate the Debye logarithm as roughly constant at a given redshift, $\Lambda(z) =\ln (2 \mu v_\text{rel} / m_D)$ since the support from the velocity distribution in the integrand is not especially broad. To determine a typical value of $v_\text{rel}$ inside the Debye log at a given redshift, we add the velocity dispersion with the bulk fluid relative velocity in quadrature to smoothly transition between epochs where one or the other is the relevant velocity scale in the problem. This logarithmic factor changes by less than 10\% over the range of redshifts we consider. {In order to set constraints with CMB data, we evaluate likelihoods varying over the normalization of the scattering rate for a given DM mass. To do so, we define the cross section 
\begin{align}
    \sigma_0 \equiv \frac{4 \pi Q^2 \alpha^2 \Lambda(z=10^7)}{\mu^2}, 
    \label{eq:sig0_def}
\end{align} 
where we have factored out the $v_{\rm rel}^{-4}$ and anchored the Debye logarithm at $z = 10^7$, since at these early times (when $v_{\rm rel} \to 1$) it is constant for the DM mass range we consider. Note this treatment differs slightly from previous studies of DM-baryon scattering, which considered DM with cold initial conditions and assumed a constant Debye logarithm~\cite{Dvorkin:2013cea,Dvorkin:2019zdi,Slatyer:2018aqg,Boddy:2018wzy}.

Performing the integrals in Eq.~\eqref{masterdrag} over the initial state momenta weighted by the phase-space distributions is subtle because the DM and baryons have different coherent bulk velocities due to baryon acoustic oscillations, even in the case where the DM only interacts with baryons gravitationally. We emphasize that this difference in bulk velocities $\vec{V}_\text{rel} =\vec{V}_\chi - \vec{V}_b $ is different from the $\vec{v}_\text{rel}$ appearing in the equations above, which is the relative velocity on a particle-by-particle interaction basis rather than in the bulk. The momentum-transfer cross section for DM-baryon scattering depends on the scattering kinematics only through $v_\text{rel}$, so the integrand is Galilean invariant provided that the phase space factors are boosted accordingly. Acceleration and Hubble damping of peculiar velocities are also Galilean invariant quantities, indicating freedom in the choice of frame for determining the DM-baryon drag. It is convenient to work in the rest frame of the baryons, where $\vec{V}_\chi = \vec{V}_\text{rel}$ and the drag coefficient $R_\chi$ can be read off from $d \vec{V}_\chi /d t$. We take the baryon momenta to be Maxwell-Boltzmann distributed so that the phase space for protons (the dominant species responsible for drag) is $f_{b,v}(p_b) =  (2 \pi / m_p T_b)^{3/2} \exp({-p_b^2 / 2 m_p T_b})$ for baryon temperature $T_b$. For the DM velocity distribution, we must boost the phase space in the bulk DM frame, $f_{\chi, v; V_\chi =0}(p_\chi)$ to the baryon frame. Taking $\vec{v}_\chi$ as the velocity of an individual DM particle in the bulk rest frame of the baryons, we then evaluate the DM phase space in the baryon frame as $f_{\chi, v; V_\chi =0}(m_\chi\abs{\vec{v}_\chi - \vec{V}_\chi })$. For the freeze-in scenarios here, we consider either a Maxwell-Boltzmann distribution, or a nonthermal distribution; the latter case is possible if the dark gauge coupling is not large, and the phase space is solely determined numerically through the methods in Ref.~\cite{Dvorkin:2019zdi}, which worked in the bulk DM rest frame at early times corresponding to $T_b\sim$~MeV temperatures.

\subsection{Gaussian DM Phase Space}
If the DM has relatively strong self-interactions, then it is possible for DM to self-thermalize prior to recombination while still evading bounds from merging clusters of galaxies~\cite{Dvorkin:2019zdi}. The DM phase space in the bulk baryon rest frame would then be $f_{\chi, v; V_\chi =0}(m_\chi\abs{\vec{v}_\chi - \vec{V}_\chi }) =  (2 \pi / m_\chi T_\chi)^{3/2} \exp({-m_\chi(\vec{v}_\chi - \vec{V}_\chi)^2 / 2 T_\chi})$ where the DM temperature $T_\chi$ can be calculated based on the second moment of the distribution $\left<p_\chi^2\right>$, which is conserved if the DM thermalizes with itself in isolation from the SM (due to conservation of energy). If the DM phase space is Gaussian, we may use the well-known techniques of Refs.~\cite{Dvorkin:2013cea,Boddy:2018wzy,Munoz:2015bca} to compute the drag force between the DM and baryon fluids. In particular, we can make a convenient change of variables to relative velocity and thermally averaged bulk velocity,  
\beq \vec{v}_\text{rel}= \vec{v}_\chi - \vec{v}_b  \quad \quad \quad \vec{v}_\text{bulk} = \frac{\frac{T_\chi}{m_\chi} \vec{v}_b+\frac{T_b}{m_p} \vec{v}_\chi}{\frac{T_\chi}{m_\chi}+ \frac{T_b}{m_p}},\label{variabletrans}\eeq
so that in the new variables the integrals over the phase space can be written as 
\beq \int  \dbar^3 p_{\chi} ~\dbar^3 p_{b} f_{\chi, v}(p_{\chi}) f_{b, v}(p_{b}) = \frac{1}{(2 \pi)^3} \int  d^3 v_\text{rel} ~d^3 v_\text{bulk} \frac{1}{\sigma_\text{rel}^3} e^{-(\vec{v}_\text{rel} - \vec{V}_\chi)^2/2\sigma_\text{rel}^2 } \frac{1}{\sigma_\text{bulk}^3} e^{-(\vec{v}_\text{bulk} - \vec{V}_\text{bulk})^2/2 \sigma_\text{bulk}^2},\eeq
where the thermal dispersions of the relative velocity and thermal bulk velocity are \beq 
\sigma_\text{rel}^2= \frac{T_\chi}{m_\chi} + \frac{T_b}{m_p} \quad \quad \quad \sigma_\text{bulk}^2 =\left( \frac{m_\chi}{T_\chi} + \frac{m_p}{T_b}\right)^{-1} \label{eq_variances} \eeq
and the mean thermal bulk velocity is 
\beq \vec{V}_\text{bulk}= \frac{T_b m_\chi \vec{V}_\chi }{T_b m_\chi +T_\chi m_p}.\eeq
With this convenient change of variables, we obtain 
\begin{align*} \frac{d \vec{V}_\chi}{dt} +  H \vec{V}_\chi &= -\frac{\sqrt{2}\rho_b X_p \sigma_0 \vec{V}_\chi}{3 \sqrt{\pi}(m_\chi+m_p)\sigma_\text{rel}^3} ~_1F_1 \left(\frac{3}{2}, \frac{5}{2},-\frac{V_\chi^2}{2 \sigma_\text{rel}^2}\right)\\
&=\frac{\rho_b X_p \sigma_0 \vec{V}_\chi}{(m_\chi+m_p)V_\chi^3}\left(\text{Erf}\left(\frac{V_\chi}{\sqrt{2}\sigma_\text{rel}}\right) -\sqrt{\frac{2}{\pi}} \frac{V_\chi}{\sigma_\text{rel}} e^{-V_\chi^2/2 \sigma_\text{rel}^2}\right). \label{gaussiandrag} \end{align*} 

\subsection{Non-Gaussian DM Phase Space}
If the DM does not self-thermalize prior to recombination, then the change of variables in the previous Subsection will not prove useful since the DM does not possess a temperature and more care must be taken with the velocity exchange rate (as was pointed out in Ref.~\cite{Ali-Haimoud:2018dvo}). Instead, we first perform the integral over initial baryon momenta. To perform this integral, we choose a coordinate system where $\vec{v}_\chi = v_\chi \hat{z}$ so that $\vec{v}_\text{rel} = (-v_b \sin \theta_b \cos \phi_b , -v_b \sin \theta_b \sin \phi_b, v_\chi - v_b \cos \theta_b)$.  
The integration proceeds as
\begin{align}  
&-\frac{\sigma_0 \rho_b X_p}{(2\pi)^3  (m_\chi + m_p)} \int \dbar^3 p_\chi \,p_b^2 d p_b\, d \Omega_b\,  f_{\chi, v; V_\chi =0}(m_\chi\abs{\vec{v}_\chi - \vec{V}_\chi }) f_{b,v}(p_b) \frac{\vec{v}_\text{rel}}{v_\text{rel}^3}  \nonumber\\
=&-\frac{\sigma_0 \rho_bX_p}{(2 \pi)^2  (m_\chi + m_p)} \int \dbar^3 p_\chi \,p_b^2 d p_b\, d \cos\theta_b\,  f_{\chi, v; V_\chi =0}(m_\chi\abs{\vec{v}_\chi - \vec{V}_\chi }) f_{b,v}(p_b) \frac{\vec{v}_\chi}{v_\text{rel}^3} (1-v_b \cos \theta_b /v_\chi ) \nonumber\\
= &- \frac{2\sigma_0 \rho_b X_p}{(2 \pi)^2  (m_\chi + m_p)} \int \dbar^3 p_\chi \, f_{\chi, v; V_\chi =0}(m_\chi\abs{\vec{v}_\chi - \vec{V}_\chi }) \frac{\vec{v}_\chi}{v_\chi^3} \int_0^{m_p v_\chi} \,p_b^2 d p_b\,  f_{b,v}(p_b) \nonumber\\
=&-\frac{2\sigma_0 \rho_b X_p}{(2 \pi)^2  (m_\chi + m_p)} \int \dbar^3 p_\chi  f_{\chi, v; V_\chi =0}(m_\chi\abs{\vec{v}_\chi - \vec{V}_\chi })\frac{\vec{v}_\chi }{v_\chi^3} \left(2 \pi^2 \text{Erf}\left( \sqrt{\frac{m_p}{2 T_b}} v_\chi\right) - \sqrt{\frac{(2 \pi)^3 m_p}{T_b}} v_\chi e^{- m_p v_\chi^2 /2 T_b} \right) .
\end{align}
Between the first and second line, only the $z$-component of the relative velocity vector had a non-zero azimuthal average; this has been expressed in terms of the direction of $\vec{v}_\chi$, making the expressions above independent of choice of coordinate system. In other words, had we chosen another coordinate system, only the projection of the baryon momentum onto the DM momentum would have survived azimuthal integration. Between the second and third line, the integration over the baryon declination angle imposed a baryon momentum restriction. Integrating over the azimuthal angle of the DM particle momentum, we find
\begin{align*} \frac{d \vec{V}_\chi}{dt} +  H \vec{V}_\chi =-\frac{2 \sigma_0 \rho_b X_p m_\chi^3 \hat{V}_\chi}{(2 \pi)^4  (m_\chi+m_p)} \int  d v_\chi d \cos \theta_\chi & f_{\chi, v; V_\chi =0}(m_\chi\abs{\vec{v}_\chi - \vec{V}_\chi }) \cos\theta_\chi\\ \times & \left(2 \pi^2 \text{Erf}\left( \sqrt{\frac{m_p}{2 T_b}} v_\chi\right) - \sqrt{\frac{(2 \pi)^3 m_p}{T_b}} v_\chi e^{- m_p v_\chi^2 /2 T_b} \right),\label{intovervx}\numberthis\end{align*}
where, as with the above cases, the only component of DM velocity surviving angular integration is the component that is projected onto the bulk relative velocity; the remaining angular variable $\theta_\chi$ is interpreted as the angle between $\vec{v}_\chi$ and $\vec{V}_\chi$.

\subsection{Treatment of bulk velocities}
Further analytic progress can be made in the non-Gaussian case if there is a hierarchy of velocities allowing for Eq.~\eqref{intovervx} to be expanded. We take the approach of Refs.~\cite{Dvorkin:2013cea, Xu:2018efh} and note that the root-mean-square (RMS) bulk velocity difference between the DM and baryon fluids is ~\cite{tseliakhovich2010relative}
\begin{align}
	V_{\rm RMS}^2 &\equiv  \langle |{\bf V}_{\chi} - {\bf V}_{b}|^2 \rangle = \int \frac{dk}{k} \Delta_\zeta^2(k) \left( \frac{ \theta_b(k,z) - \theta_\chi(k,z)}{k} \right)^2 \\
    & \simeq \begin{cases}
         10^{-8}, & z > 10^3 \\
         10^{-8} \left( \tfrac{1+z}{10^3} \right)^2, & z \le 10^3,
    \end{cases}
    \label{eq:VRMS}
\end{align}
where $\Delta_\zeta^2(k)$ is the dimensionless power spectrum of primordial perturbations to the curvature $\zeta$, with $\Delta_\zeta^2(k) \approx 2.4 \times 10^{-9}$. Here we have made the same approximation as in previous works~\cite{Dvorkin:2013cea,Dvorkin:2019zdi,Slatyer:2018aqg} of integrating over all $k$ modes for the bulk velocity, despite the fact that we are considering evolution of individual $k$ modes. Ref.~\cite{Boddy:2018wzy} presented a prescription to account for the $k$-dependence of the bulk velocity, and found that the limits on DM scattering are very similar regardless of whether or not the $k$-dependence is included (including $k$-dependence made their constraints very slightly stronger than ignoring it). Because the impact of $k$-dependence on the limits is small and because the prescription of Ref.~\cite{Boddy:2018wzy} is strictly speaking only applicable when both DM and baryons have a Gaussian velocity distribution, we ignore those effects in this work. In Fig.~\ref{bulkvsthermal} we compare this typical relative bulk velocity to the RMS velocities of DM and baryons in the rest frames of their respective fluids. 
\begin{figure}[t]
    \centering
    \includegraphics[width=0.5\textwidth]{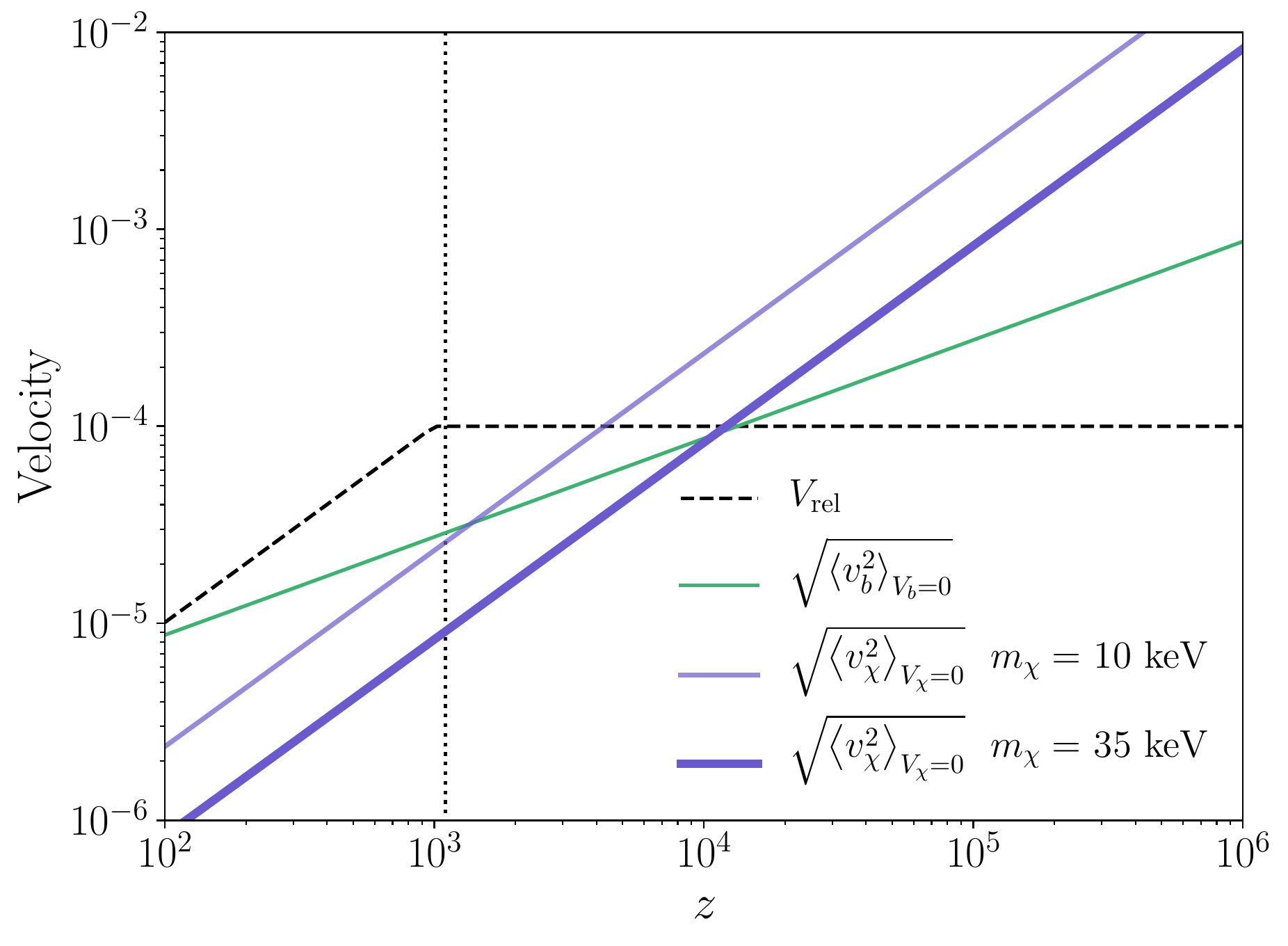}\includegraphics[width=0.5\textwidth]{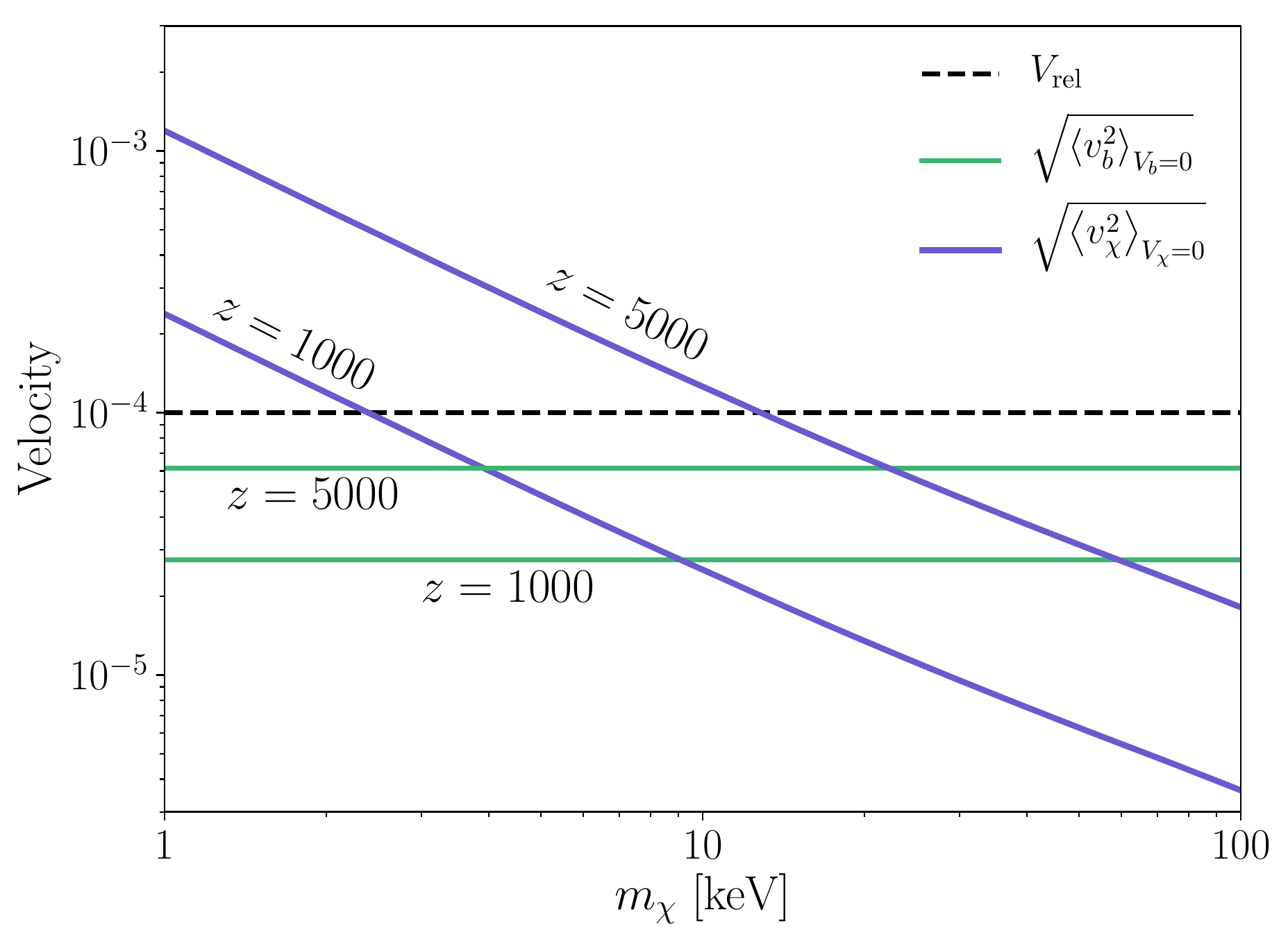}
    \caption{Root-mean-square peculiar velocities of the DM and baryon fluids (in units of $c$), evaluated in the rest frames of those respective fluids. For freeze-in, this quantity is the same whether the distribution is thermal or nonthermal, as the second moment of the phase-space distribution is conserved during thermalization within the DM sector. \emph{Left:} As this sector is secluded from the photons and baryons, the typical velocity scales like $(1+z)$. Meanwhile, the typical baryon velocity scales like $(1+z)^{1/2}$ because the nonrelativistic baryons are in kinetic equilibrium with the photons which have a $(1+z)$ temperature scaling. Also shown is the typical bulk velocity between the two fluids, $V_\text{rel}$, that is induced by baryon acoustic oscillations. This bulk velocity drops off shortly after recombination (vertical dotted line). \emph{Right:} The same velocities shown at $z=1000$ and $z=5000$ as a function of DM mass. Lighter DM masses have relatively high velocities close to recombination, meaning that the freeze-in initial conditions affect the relative velocity and the drag force most for these masses.} 
    \label{bulkvsthermal}
\end{figure}
Note that this RMS DM speed is the same regardless of whether the DM has thermalized or not, since this is precisely the moment of the phase-space distribution that must be conserved if the DM self-thermalizes while nonrelativistic, due to conservation of energy within this secluded DM fluid. 

As shown in Fig.~\ref{bulkvsthermal}, the typical DM speed is much larger than the relative fluid velocities for much of the expansion history prior to recombination, so we can treat $V_\chi/v_\chi$ as a small parameter during these epochs. Expanding $\abs{\vec{v}_\chi - \vec{V}_\chi} \approx v_\chi (1 - V_\chi \cos \theta_\chi / v_\chi)$, the phase-space distribution becomes 
\beq f_{\chi,v}(m_\chi\abs{\vec{v}_\chi - \vec{V}_\chi}) \approx f_{\chi,v}(m_\chi v_\chi) + f'_
{\chi,v}(m_\chi v_\chi)(\abs{\vec{v}_\chi - \vec{V}_\chi}-v_\chi)m_\chi \approx f_{\chi,v}(m_\chi v_\chi) - f'_{\chi,v}(m_\chi v_\chi) m_\chi V_\chi \cos \theta_\chi \eeq at leading order in $V_\chi$. Only the term that is linear in $\cos \theta_\chi$ will survive angular integration because of the additional factor of $\cos\theta_\chi$ appearing in Eq.~\eqref{intovervx}.
Integrating by parts and dropping surface terms, we find 
\begin{align} 
\lim_{V_\text{rel}\rightarrow 0} R_\chi &= \frac{4 \sigma_0 \rho_b X_p m_\chi^3 a}{3 (2 \pi)^4  (m_\chi+m_p)}
\int  d v_\chi f_{\chi, v; V_\chi=0}(m_\chi v_\chi)   \left( \frac{2 \pi m_p}{T_b}\right)^{3/2} v_\chi^2 e^{-\frac{m_p v_\chi^2}{2 T_b}}. \label{lowVexpansion}
\end{align}
For the case of a nonthermal DM velocity distribution, the remaining integral must be performed numerically given the distribution functions from the two freeze-in channels~\cite{Dvorkin:2019zdi}. In the case of a Gaussian DM velocity distribution (if the DM can thermalize in its own sector significantly before recombination), Eq.~\eqref{lowVexpansion} reduces to Eq.~(14) of Ref.~\cite{Dvorkin:2013cea} which first derived this result, which is also the low-$V_\chi$ limit of Eq.~\eqref{gaussiandrag}. This expansion shows explicitly how the drag rate at early times is sensitive to the full DM phase space distribution function, and accounts for the differences between freeze-in DM that has thermalized and freeze-in DM that has not thermalized. We have verified that this low-$V_\chi$ approximation works very well at early times and matches the full drag force computed with Eq.~\eqref{intovervx}. 

At later times, the bulk velocity between the DM and baryon fluids is larger than the typical DM velocity in the rest frame of the DM fluid, $V_\text{rel} \gg \sqrt{\left< v_\chi^2\right>_{V_\chi=0}}\equiv v_{\chi}^\text{RMS}$. Since the width of the peak of the phase-space distribution $\sim v_{\chi}^\text{RMS}$ is very narrow compared to the relevant velocity scale in the problem $V_\text{rel}$, the exact shape of the distribution should not affect the drag force. For the integral to have any support, there must be a fine cancellation between $\vec{v}_\chi$ and $\vec{V}_\chi$ in the argument of the phase-space distribution in Eq.~\eqref{intovervx}, meaning that the phase space essentially acts like a delta function. In Fig.~\ref{dragrate}, we indeed see that the drag force does not depend on whether the distribution has thermalized when $V_\text{rel} > v_{\chi}^\text{RMS}$ around the time of recombination. 

\begin{figure}[t]
    \centering
    \includegraphics[width=0.6\textwidth]{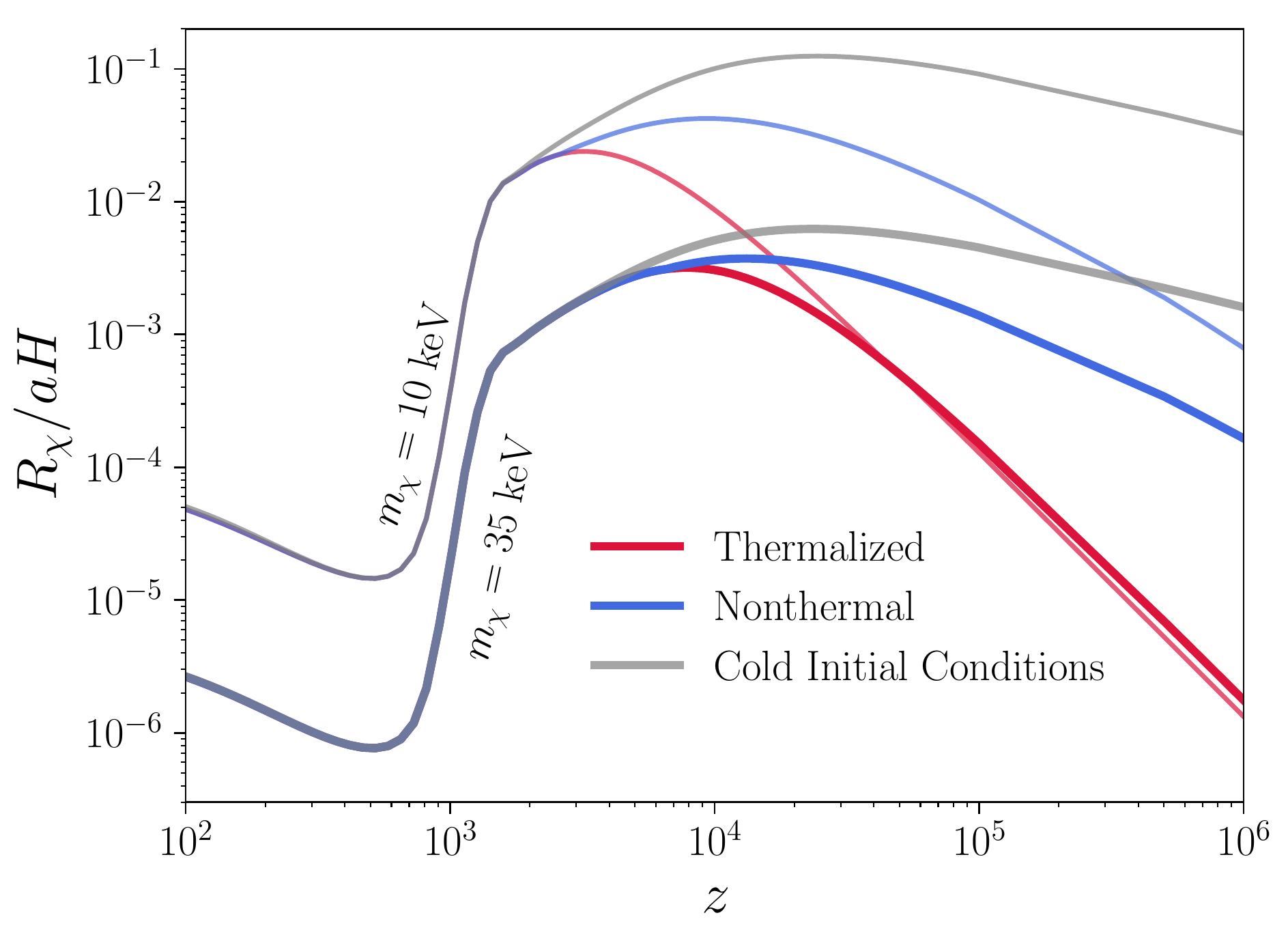}
    \caption{Drag rate between DM and baryon fluids in Hubble units, assuming the DM-baryon couplings required for the freeze-in mechanism to make all the DM. Thick lines correspond to $m_\chi=35$~keV while thin ones correspond to $m_\chi=10$~keV. The colors correspond to differences in the DM phase space. Since the scattering cross section scales like $v_\text{rel}^{-4}$, the drag is largest if the DM has cold initial conditions. The nonthermal freeze-in phase space yields a larger drag rate than the thermalized one since the distribution is peaked at the lower momenta with a long tail. At times near recombination, when the bulk velocity $V_\text{rel}$ is larger than the DM velocity dispersion, the phase-space distribution becomes irrelevant in determining the drag between the two fluids and the thermal and nonthermal scenarios yield the same drag rate as in the case of cold initial conditions.}
    \label{dragrate}
\end{figure}

\section{Heat Transfer between DM and Baryon Fluids}
Scattering between DM and baryons can transfer heat between the two fluids, which is captured by the second moment of the Boltzmann hierarchy, \begin{align}
\label{eq:tempev}
	\dot T_\chi^\text{eff} &= -2 \frac{\dot a}{a} T_\chi^\text{eff}    +   \frac{2 a}{3} \frac{dQ_\chi}{dt}, \\
	\dot T_b &= -2 \frac{\dot a}{a} T_b    +   \frac{2 \mu_b}{m_e} R_\gamma (T_\gamma - T_b)  +   \frac{2 a}{3} \frac{dQ_b}{dt}.
\label{eq:barytempev}
\end{align} 
Here, $T_\chi^\text{eff}$ is the effective DM temperature, $3 T_\chi^\text{eff} = m_\chi \left(v_\chi^\text{RMS}\right)^2$, which is the same regardless of whether DM self-scattering has thermalized the DM, $\mu_b$ is the mean molecular weight of the baryons, and $Q_i$ is the heat transferred to species $i$. Note again that overdots denote derivatives with respect to conformal time. The rate of heat transfer to the DM over cosmic time is 
\begin{align*} \frac{d Q_\chi}{dt} &= -\frac{1}{2\rho_\chi} \int \frac{ \dbar^3 p_{\chi,i}}{2 E_{\chi,i }}\frac{ \dbar^3 p_{b,i}}{2 E_{b,i}}  \frac{ \dbar^3 p_{\chi,f}}{2 E_{\chi,f}}  \frac{ \dbar^3 p_{b,f}}{2 E_{b,f}} \sum_\text{dof} \abs{\mathcal{M}}^2 (2 \pi)^4 \delta^{(4)}\left(p_{\chi,i}+p_{b,i}- p_{\chi,f} - p_{b,f}\right) \vec{p}^2_{\chi,i} \left(f_{\chi,i} f_{b,i} - f_{\chi,f} f_{b,f}\right)\\
&= -\frac{1}{2\rho_\chi} \int \frac{ \dbar^3 p_{\chi,i}}{2E_{\chi,i }}\frac{ \dbar^3 p_{b,i}}{2 E_{b,i}}  \frac{ \dbar^3 p_{\chi,f}}{2 E_{\chi,f}}  \frac{ \dbar^3 p_{b,f}}{2 E_{b,f}} \sum_\text{dof} \abs{\mathcal{M}}^2 (2 \pi)^4 \delta^{(4)}\left(p_{\chi,i}+p_{b,i}- p_{\chi,f} - p_{b,f}\right) (\vec{p}^2_{\chi,i} - \vec{p}^2_{\chi,f})  f_{\chi,i} f_{b,i} , \numberthis
\end{align*}
where the same reasoning as below Eq.\eqref{eq:rawcoll} applies between the first and second line. The energy transferred to the DM in a single collision can be rewritten as \beq \Delta E_\chi = \frac{1}{2 m_\chi}\left(\vec{p}^2_{\chi,f}-\vec{p}^2_{\chi,i} \right) = \frac{1}{2 m_\chi}\left( 2\vec{p}_{\chi,i}+ \Delta \vec{p}_\chi \right)\cdot \Delta \vec{p}_\chi = \left( \frac{\vec{p}_{\chi,i}+\vec{p}_{b,i}}{m_\chi + m_p} \right) \cdot \Delta \vec{p}_\chi \equiv \vec{v}_\text{CM} \cdot \Delta \vec{p}_\chi, \eeq
where the third equality can be shown using conservation of energy and where $\vec{v}_\text{CM}$ is the boost required to go from some arbitrary frame to the center-of-mass frame. Writing the energy transfer in this way means that much of the same reasoning ports over from the discussion above Eq.~\eqref{finalstates} because $\vec{v}_\text{CM}$ only depends on initial-state momenta and we computed the integral weighted by $\Delta \vec{p}_\chi $ over the final-state momenta in Eq.~\eqref{finalstates}. The collisional heat transfer rate is thus 
\beq 
\frac{d Q_\chi}{dt}=-\frac{\rho_b X_p m_\chi \sigma_0}{(m_\chi + m_p)}\int \dbar^3 p_{\chi} ~\dbar^3 p_{b} f_{\chi, v}(p_{\chi}) f_{b, v}(p_{b})\, \frac{\vec{v}_\text{CM}\cdot \vec{v}_\text{rel} }{v_\text{rel}^3}.\label{heatgeneral}\eeq

\begin{figure}[t]
    \centering
    \includegraphics[width=0.6\textwidth]{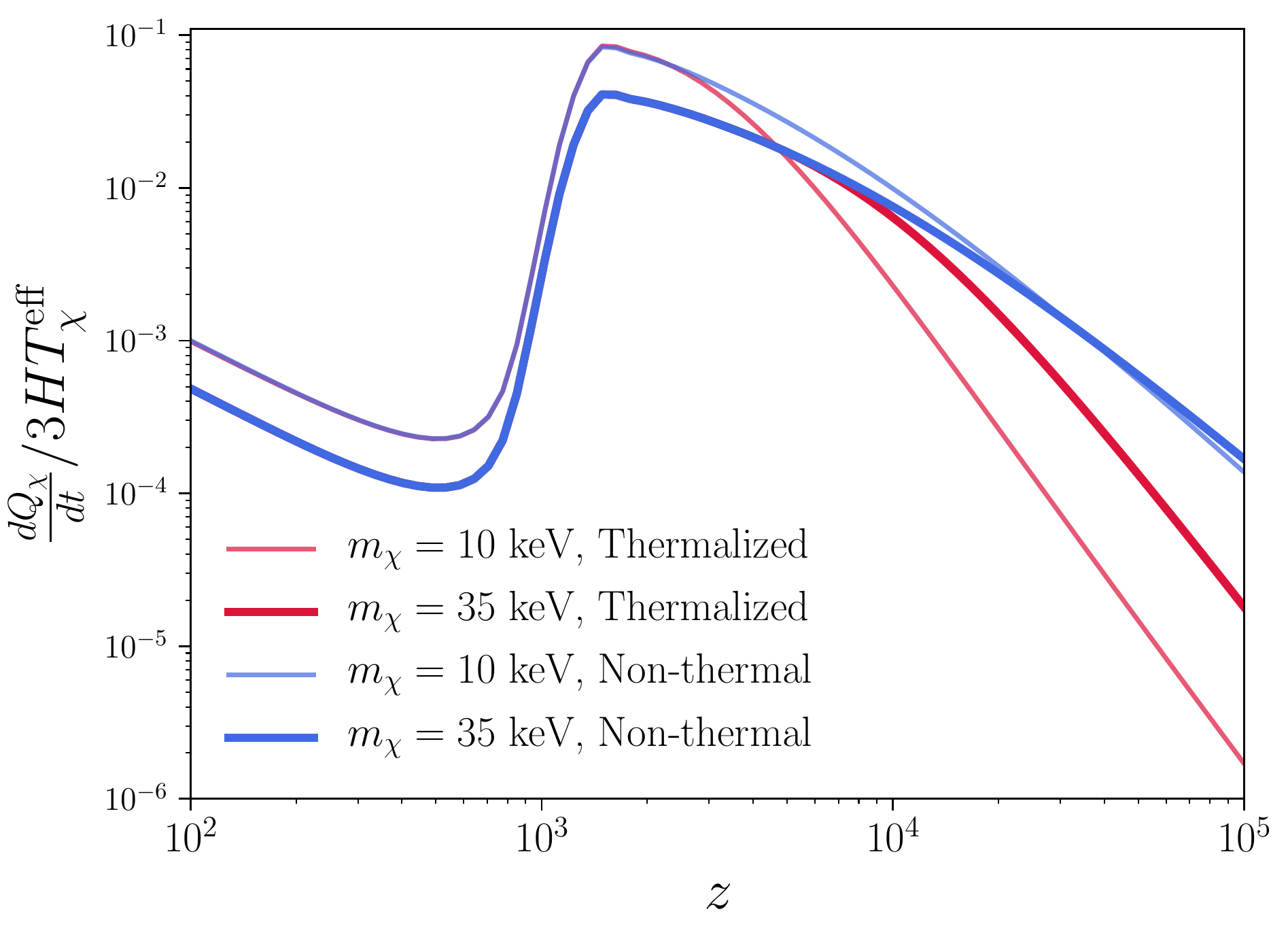}
\caption{The collisional DM heating rate in the DM rest frame compared to the cooling rate due to Hubble expansion. As with the drag rate, at early times when the DM velocity dispersion is larger than the bulk fluid velocity, the heating rate for the nonthermal DM phase-space distribution is higher than for the thermal one. Later when the bulk velocity becomes the largest velocity scale in the problem, the DM phase space becomes irrelevant and the heating rate is the same for the two kinds of distributions. The collisional heating rate is always highly subdominant to the expansion cooling rate whether the DM has thermalized or not. } 
    \label{heating}
\end{figure}

The heat transfer rate is not a Galilean invariant quantity; in particular, $\vec{v}_\text{CM}$ appearing in the energy transfer of a single scattering event is frame dependent. We choose to evaluate the collisional DM heating rate in the bulk rest frame of the DM fluid and will compare this heating to $\left<p_\chi^2\right>/2 m_\chi$ also evaluated in the bulk rest frame. This circumvents the need to think about the DM kinetic energy due to bulk motions, which would be present regardless of any exotic DM microphysics. We therefore boost the baryon phase space to that frame, $f_{b, v}(p_{b}) = (2 \pi / m_p T_b)^{3/2} \exp({-(\vec{p}_b - m_p \vec{V}_b)^2 / 2 m_p T_b})$. We also note that when considering the drag force, the total momentum (net force) is conserved; however, for energy exchange between the two fluids, the total thermal energies may not be conserved since there is also energy in the bulk flow~\cite{Munoz:2015bca}.
\subsection{Gaussian DM Phase Space}
If the DM phase space distribution is Gaussian, then we can make the same change of variables from DM and baryon velocities to relative and thermally averaged bulk velocities as in Eq.~\eqref{variabletrans}. Performing this transformation in Eq.~\eqref{heatgeneral} straightforwardly yields 
\beq \frac{d Q_\chi}{dt}= \frac{\rho_b X_p m_\chi \sigma_0}{(m_\chi + m_p)^2 \sigma_\text{rel}^3}
\left[\sqrt{\frac{2}{\pi}}(T_b - T_\chi)e^{-V_b^2/2\sigma_\text{rel}^2 } + \frac{m_p \sigma_\text{rel}^3}{V_b} \left(\text{Erf}\left(\frac{V_b}{\sqrt{2}\sigma_\text{rel}}\right) -\sqrt{\frac{2}{\pi}} \frac{V_b}{\sigma_\text{rel}} e^{-V_b^2/2 \sigma_\text{rel}^2}\right) \right],\eeq
where $\sigma_\text{rel}$ is defined in Eq.~\eqref{eq_variances}. This heating rate for freeze-in initial conditions is shown in Fig.~\ref{heating} in red for the case where DM thermalizes.

\subsection{Non-Gaussian DM Phase space}
Since the DM phase space is isotropic in the bulk rest frame of the DM fluid, we can analytically integrate over the DM solid angle
\beq \int d \Omega_\chi \frac{\vec{v}_\text{CM}\cdot \vec{v}_\text{rel} }{v_\text{rel}^3} = 4 \pi \mu \times 
\begin{cases} 
      \frac{1}{m_p v_\chi} & v_b<v_\chi \\
      -\frac{1}{m_\chi v_b} & v_b>v_\chi   
   \end{cases}
 \eeq
We can also analytically integrate over the baryon solid angle,
\beq \int d \Omega_b f_{b,v}(p_b) = 2 \sqrt{\frac{(2 \pi)^5}{m_p^5 T_b}}\frac{\, e^{-\frac{m_p(v_b^2 + V_b^2)}{2 T_b}}\sinh{\left(\frac{m_p v_b V_b}{T_b}\right)}}{v_b V_b}.\eeq Upon integrating over the baryon velocity, the collisional heating rate is 
\begin{align*} \frac{m_\chi^4 \rho_b X_p \sigma_0 }{4\sqrt{\pi^5}(m_p+m_\chi)^2 V_b } &\int  d v_\chi v_\chi 
f_{\chi,v; V_\chi=0}(m_\chi v_\chi)  \Bigg[ 2 \sqrt{\frac{2 T_b}{m_p}} m_\chi e^{-m_p (v_\chi^2 + V_b^2)/2 T_b} \sinh \left( \frac{m_p V_b v_\chi}{T_b}\right) \numberthis\label{fullheat}\\&+ \sqrt{\pi} \left((m_p v_\chi+m_\chi V_b )\text{ erf} \left(\sqrt{\frac{m_p}{2 T_b}}(V_b-v_\chi) \right) +(m_p v_\chi-m_\chi V_b )\text{ erf} \left(\sqrt{\frac{m_p}{2 T_b}}(V_b+v_\chi) \right)\right)\Bigg].\end{align*} For freeze-in initial conditions in the case where DM does not self-thermalize, the heating rate is shown in Fig.~\ref{heating} in blue.

\subsection{Effects of DM-baryon heat exchange}
For the range of DM masses we consider in this work, the effects of baryons heating the DM are entirely negligible in terms of the observable effects we consider. The heating rate is peaked close to $z\lesssim 10^4$ and we have determined that this heating can cause the DM effective temperature to increase by $\sim 10 \%$ by recombination. However, we have checked that this does not affect the DM-baryon drag effect at a level that we can constrain, since at these redshifts the bulk fluid motions dominate over the DM velocity dispersion from the effective temperature. Moreover, this slight heating does not affect structure formation, since the DM is non-relativistic at this epoch and heating it slightly is not enough to suppress the growth of small-scale density perturbations (the suppression of small-scale structure from the DM velocity is imprinted at much earlier times when the DM is moving much more quickly). We have also checked that the converse effect of DM cooling the baryons is negligible due to the tight coupling between the photons and baryons. We find that the change to the baryon temperature is less than one part in $10^8$.

\section{CMB Constraints from \emph{Planck} and Fisher forecasts for CMB-S4 \label{sec:planck}}

We run a Markov Chain Monte Carlo (MCMC) likelihood analysis using the publicly available \emph{Planck} 2018 likelihoods for the CMB temperature and polarization at small scales (TTTEEE) and large scales (lowl+lowE) \cite{Aghanim:2019ame}. We also include the lensing power spectrum~\cite{Aghanim:2018oex}.
We use the code CosmoMC \cite{Lewis:2002ah} for this analysis. For each DM mass, we include in CAMB the DM-baryon drag calculated for that mass, and leave as a free parameter the scattering cross section $\sigma_0$ defined in Eq.~\eqref{eq:sig0_def}. As discussed above, we take $\sigma_0$ to be normalized at $z=10^7$ for freeze-in, which differs somewhat from the normalization with previous studies considering DM with purely cold phase space and neglecting the Debye log. The cosmological parameters we include in our analysis are $\sigma_0$ and the standard $\Lambda$CDM parameters: baryon density $\Omega_b h^2$, DM density $\Omega_c h^2$, the optical depth to reionization $\tau$, the effective angular scale of the sound horizon at recombination $\theta_{\rm MC}$, and the amplitude and tilt of scalar perturbations $\ln A_s$ and $n_s$. We assume flat and uninformative priors on these parameters. We test the convergence of the samples by applying a Gelman-Rubin criterion \cite{gelman1992} of $R-1<0.01$ across four chains. Note that while we have included lensing power spectra, it adds only a negligible amount of significant constraining power.

In Fig. \ref{fig:nm4_2D_15keV} we show the 2-dimension posterior distributions for the standard $\Lambda$CDM cosmological parameters in addition to the scattering amplitude $\sigma_0$, for a fixed mass of $m_\chi=15$ keV. There is a correlation of DM-baryon scattering with $n_s$, and to a lesser extent with $\theta_{\rm MC}$. Both of these correlations are stronger for the nonthermal case, resulting in a weaker constraint on $\sigma_0$. This explains the counterintuitive result that the bound on $\sigma_0$ is weaker in the nonthermal case, even though the predicted change to the CMB power spectra is larger. 

With the resulting 95\% CL limits on $\sigma_0$ for different DM masses, we can find where it intersects the freeze-in prediction for $\sigma_0$ and thus obtain a lower limit on freeze-in mass. We show in Fig.~\ref{fig:allsigmabounds} the \emph{Planck} limits on $\sigma_0$ for nonthermal freeze-in DM phase space and a thermalized phase space, at masses of 15 keV and 20 keV. The gray line is the predicted $\sigma_0$ if freeze-in explains 100\% of the DM relic abundance. Interpolating between the \emph{Planck} limits, we find lower bounds of $m_\chi > 18.5$ keV for the nonthermal case and $m_\chi > 19.3$ keV for a thermalized phase space.

\begin{figure*}[th]
\includegraphics[width = \linewidth]{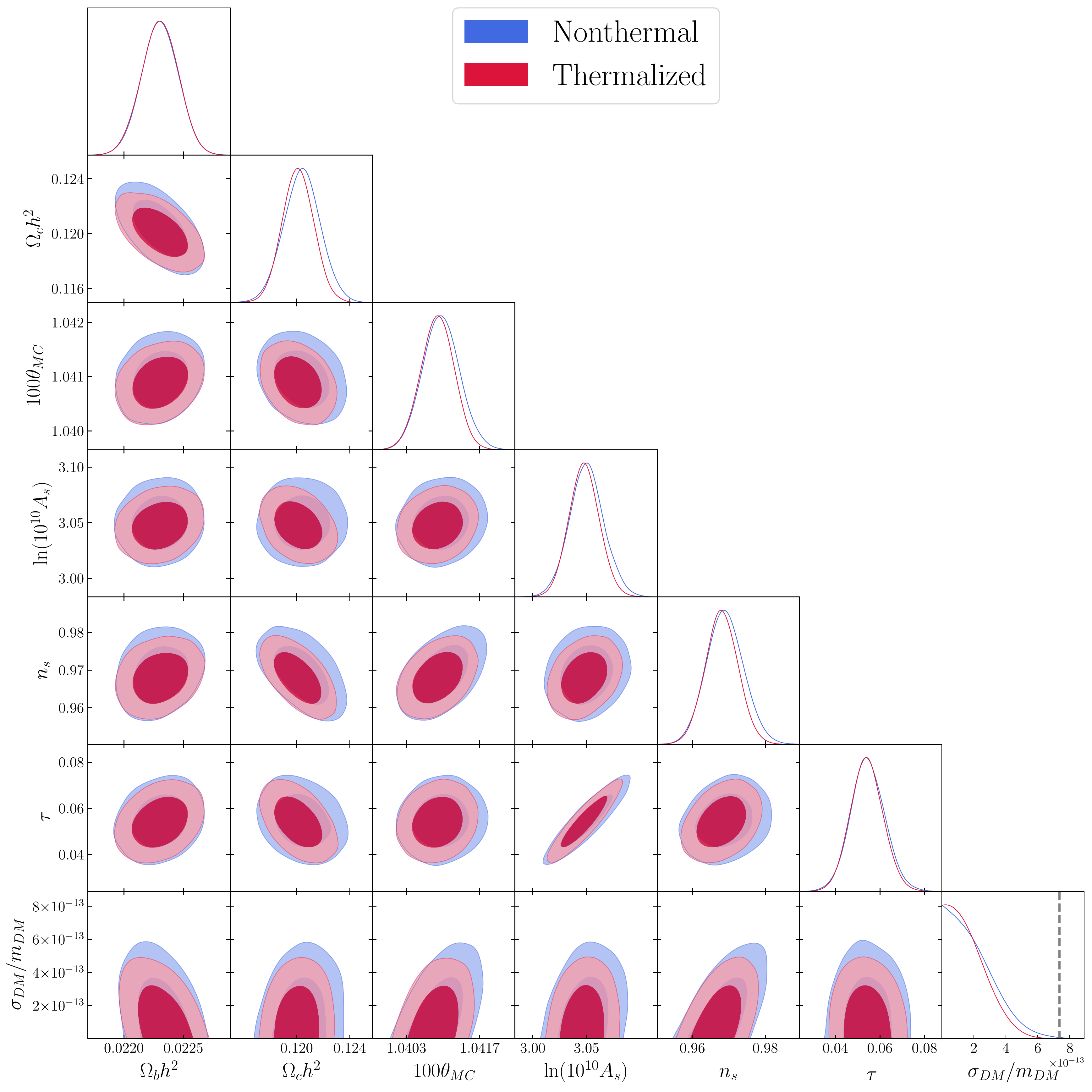}
\caption{Using the publicly available {\it Planck} 2018 likelihoods, we obtain 2-dimensional posterior distributions for the six $\Lambda$CDM cosmological parameters plus the normalization of the DM-baryon scattering rate, parameterized here as $\sigma_{\rm DM}/m_{\rm DM} = \sigma_0/m_\chi$ in units of cm$^2$/g. The DM mass is fixed at 15 keV, and the blue (red) contours are for nonthermal (thermalized) phase space. The vertical dashed line corresponds to the value of $\sigma_0/m_\chi$ appropriate for freeze-in at this DM mass.
\label{fig:nm4_2D_15keV}
}
\end{figure*}

\begin{figure}[t]
    \centering
    \includegraphics[width=0.6\textwidth]{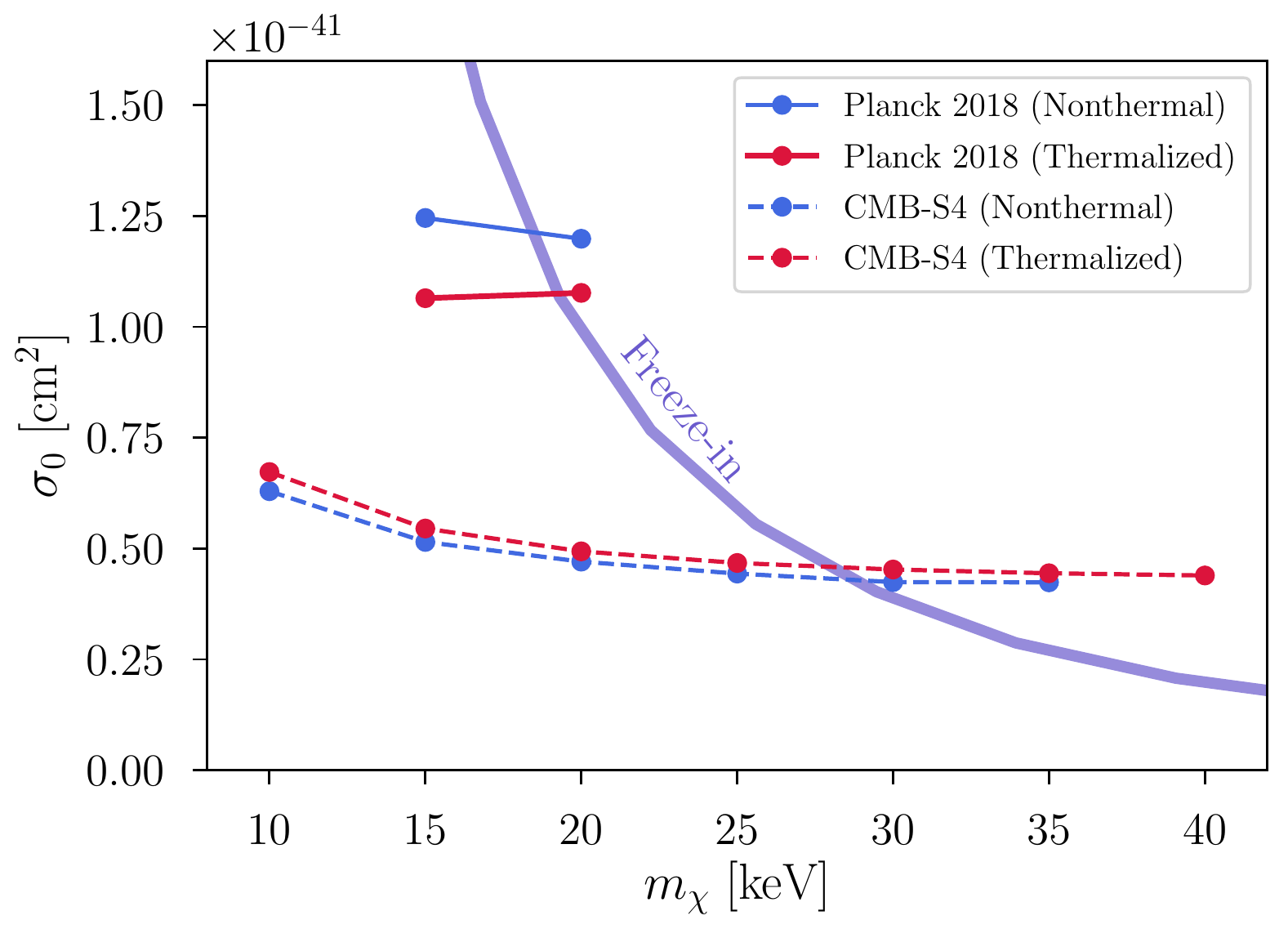}
    \caption{95\% CL upper limits on $\sigma_0$ (as defined in Eq.~\eqref{eq:sig0_def}, and normalized to the Debye logarithm at $z=10^7$) from \emph{Planck} 2018, and projected 95\% CL sensitivity with an experiment like CMB-S4 with lensing. The bounds on $\sigma_0$ are translated to lower bounds on $m_\chi$ by finding the intersection with the predicted cross section for freeze-in. \label{fig:allsigmabounds} }
\end{figure}

To forecast the sensitivity of a future CMB-S4 experiment, we use the Fisher formalism (for details, see for example Ref.~\cite{Wu:2014hta}). The Fisher matrix is given by
\begin{align}
    F_{ij} = \sum_\ell \frac{2 \ell + 1}{2} f_{\rm sky} {\rm Tr} \left( {\mathbfcal{C}}^{-1}_\ell(\arrowedvec{\theta} \,)\frac{\partial \mathbfcal{C}_\ell}{d \theta_i} {\mathbfcal{C}}^{-1}_\ell (\arrowedvec{\theta} \,) \frac{\partial \mathbfcal{C}_\ell}{d \theta_j} \right),
\end{align}
where $ \arrowedvec{\theta}$ is a vector of the fiducial six $\Lambda$CDM parameters along with the DM-baryon scattering cross section $\sigma_0$, defined in Eq.~\eqref{eq:sig0_def}. In this case for the $\Lambda$CDM parameters, we use ($\Omega_b h^2$, $\Omega_c h^2$, $\tau$, $H_0$, $\ln A_s$, $n_s$). The covariances are given by
\begin{align}
   \mathbfcal{C}_\ell = \begin{pmatrix}
C_\ell^{TT} + N_\ell^{TT} & C_\ell^{TE} & 0\\
C_\ell^{TE}  & C_\ell^{EE} + N_\ell^{EE} & 0 \\
0 & 0 & C_\ell^{dd} + N_\ell^{dd} 
\end{pmatrix}.
\end{align}
We do not include $C_\ell^{Ed}$ and $C_\ell^{Td}$ in the default analysis because they have a negligible effect on the constraints. We assume a fractional sky coverage of $f_{\rm sky} = 0.4$, and for the $\ell$ range we take $\ell_{\rm min} = 30$ with $\ell_{\rm max} = 5000$ for temperature and polarization (except for $TT$ where we assume foregrounds limit us to $\ell_{\rm max} = 3000$) and $\ell_{\rm max} = 2500$ for lensing. 
For temperature and polarization, the noise is given by
\begin{align}
    N_\ell^{TT,EE} = s_{TT,EE}^2 \exp\left( \ell (\ell + 1) \frac{\theta_{\rm FWHM}^2}{8 \ln 2} \right)
\end{align}
with $s_{TT} = 1 \mu$K-arcmin, $s_{EE} = \sqrt{2} \mu$K-arcmin and the beam resolution $\theta_{\rm FWHM} = 1$ arcmin. The lensing noise can be obtained from a procedure of iterative delensing using E-modes and B-modes \cite{Abazajian:2016yjj}.

 Since we are not adding information from large-scale polarization, we add a Gaussian prior on the optical depth of $\sigma(\tau)=0.01$, which follows the prescription used in the CMB-S4 Science Book \cite{Abazajian:2016yjj}. This amounts to taking $F_{\tau \tau} \to F_{\tau \tau} +  1/(0.01)^2$. For each parameter, the $1\sigma$ uncertainty marginalizing over all other parameters is then given by $\sigma_i = \sqrt{{\bf F}^{-1}}_{ii}$. To illustrate where the information for the constraints is coming from, in Fig.~\ref{fig:mX20_significances} we plot $\delta C_\ell^{XX}/\sigma_\ell^{XX}$ for DM-baryon scattering, where
\begin{align}
    \sigma_\ell^{XX} \equiv \sqrt{\frac{1}{f_{\rm sky}}\frac{2}{2 \ell + 1}} (C_\ell^{XX} + N_\ell^{XX})
\end{align}
for $XX = TT, EE, dd$. The quantity $\sigma_\ell^{XX}$ gives an estimate of the error as a function of $\ell$, and so $\delta C_\ell^{XX}/\sigma_\ell^{XX}$ gives an estimate of the relative significance of the various power spectra as a function of $\ell$. From this, we see that the constraints on DM-baryon drag primarily come from the suppression of the power spectra at high $\ell$.

The CMB-S4 $2\sigma$ forecasts for the DM-baryon scattering $\sigma_0$ are shown in Figs.~\ref{fig:allsigmabounds},\ref{fig:S4bounds} as a function of DM mass. As before, we interpolate the bounds in $m_\chi$ and find the intersection with the freeze-in line in Fig.~\ref{fig:allsigmabounds} to obtain a sensitivity to freeze-in mass. In Fig.~\ref{fig:S4bounds}, we further explore the mild degeneracy with $n_s$ that can already be seen in the ${\emph Planck}$ results. This degeneracy leads to larger uncertainties for $n_s$ in the CMB-S4 forecast as well. Thus, despite the fact that the significance for the nonthermal and cold initial conditions (cold ICs) cases appears to be larger in Fig.~\ref{fig:mX20_significances}, the degeneracy with $n_s$ leads to weaker bounds on $\sigma_0$ in the nonthermal case. In the case with cold ICs, the result for $\sigma_0$ is still stronger due to the much larger effect on $\delta C_\ell$. Note that for cold ICs, we have normalized $\sigma_0$ such that the drag at $z = 10^3$ is the same as for freeze-in with the same cross section. This is because the Debye logarithm at $z=10^7$ can behave quite differently depending on the choice of cold initial conditions, whereas the effect on the CMB is dominated by redshifts of $z \approx 10^3 - 10^4$.
The bottom panel of Fig.~\ref{fig:S4bounds} shows that with the inclusion of the lensing power spectra, the effect of the degeneracy with $n_s$ can be ameliorated for the nonthermal case. Then the projected bound is stronger than for a thermalized phase space.

\begin{figure*}[h!]
\includegraphics[width = \linewidth]{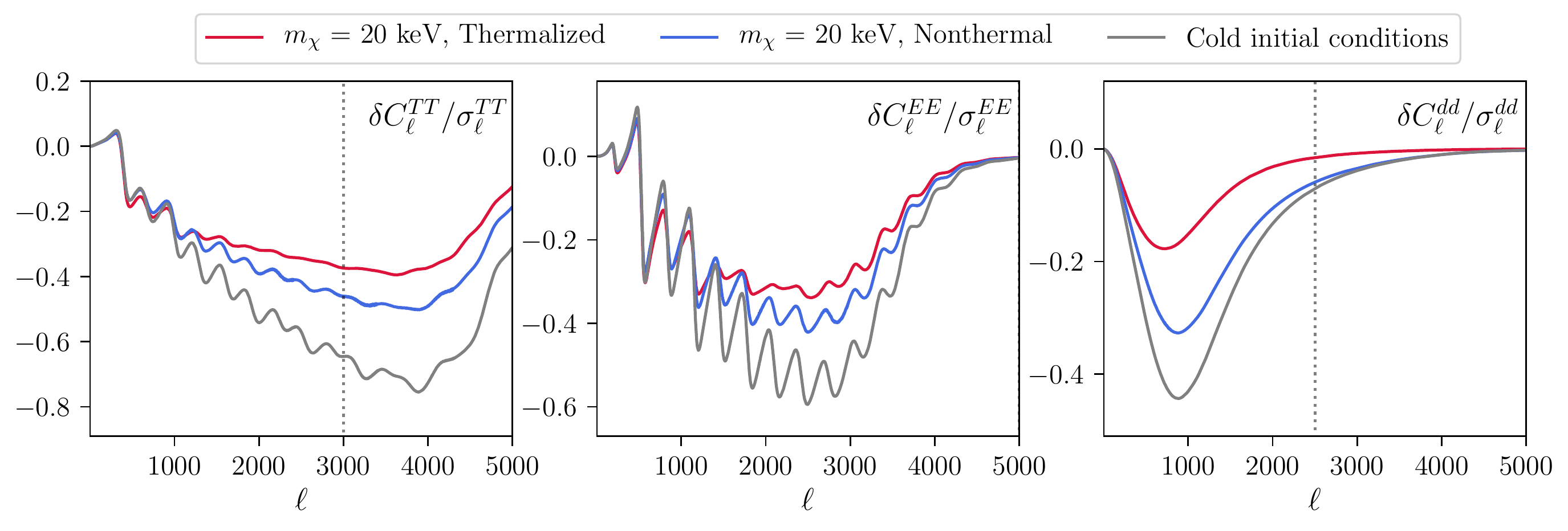}
\vspace{-0.8cm}
\caption{
Uncertainty-weighted difference in the power spectra for freeze-in as a function of $\ell$ for temperature, polarization, and lensing, for an experiment like CMB-S4. The dotted vertical lines indicate $\ell_{\rm max} = 3000, 5000$ and 2500 for temperature, polarization, and lensing, respectively. The cross section is fixed at $\sigma_0 = 10^{-41}$ cm$^2$ for freeze-in at $m_\chi = 20$ keV. For comparison, we also show the effect for DM with the same cross section and cold initial conditions.}
\label{fig:mX20_significances}
\end{figure*}

\begin{figure}[h!]
    \centering
    \includegraphics[width=0.9\textwidth]{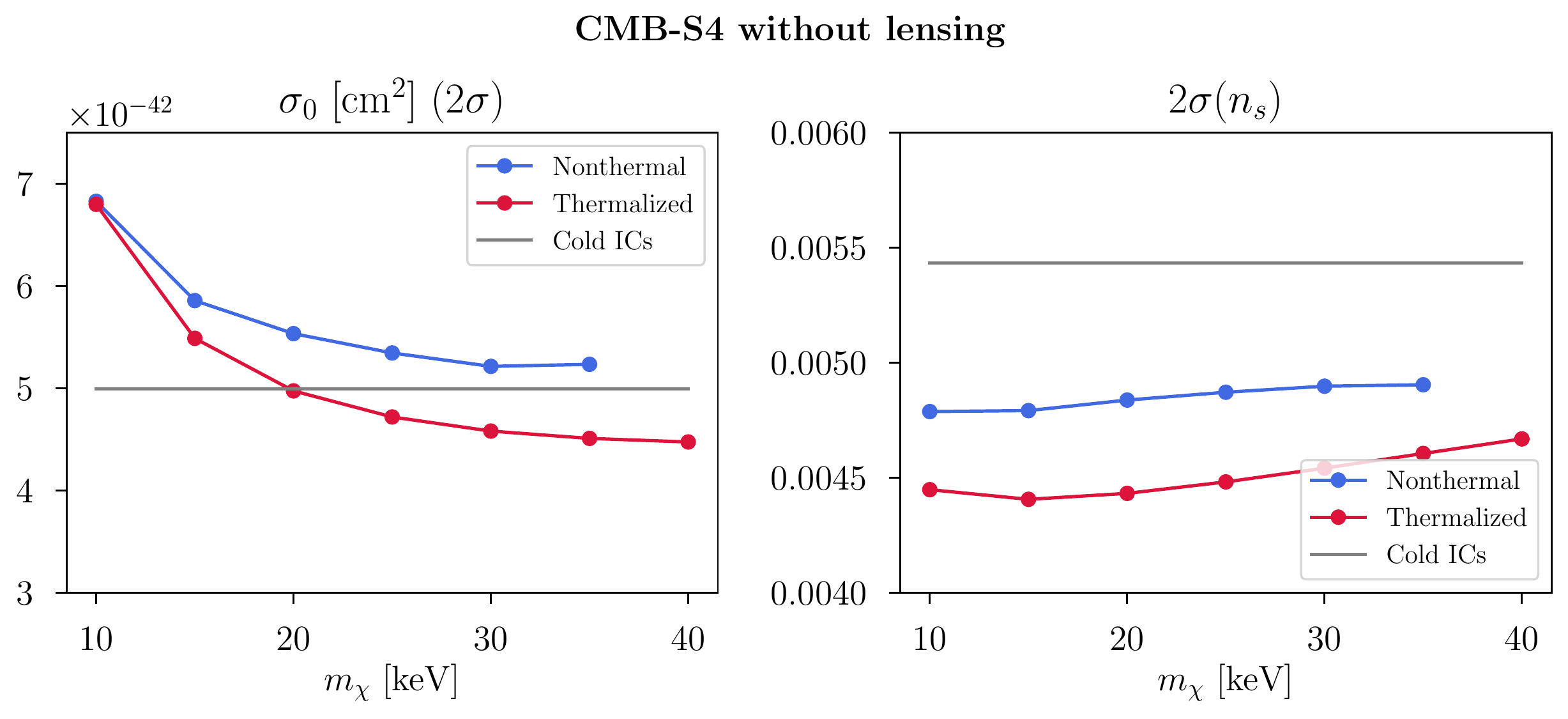}
    \includegraphics[width=0.9\textwidth]{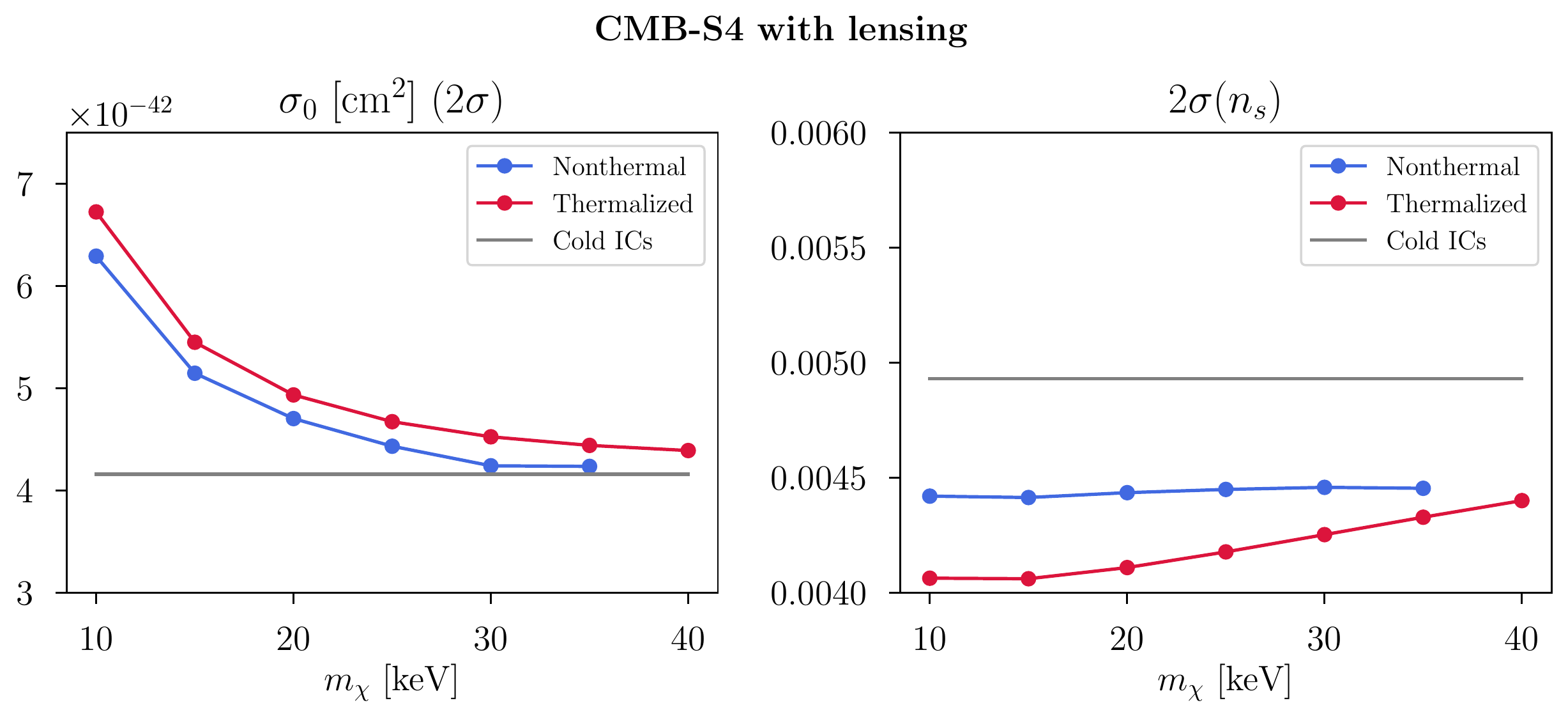}
    \vspace{-0.5cm}
    \caption{({\bf Top}) With only temperature and polarization power spectra, the sensitivity of CMB-S4 to the thermalized case is better than the nonthermal case and a DM candidate with cold initial conditions (cold ICs). Although the change to the power spectra is larger in the latter cases (see Fig.~\ref{fig:mX20_significances}), there is also a slightly larger correlation with the effect of changing $n_s$. The right panel shows how the $2\sigma$ sensitivity to $n_s$ is also significantly weaker for the nonthermal case and cold initial conditions. For all other $\Lambda$CDM parameters, the CMB-S4 sensitivity is nearly identical for all three thermal histories. ({\bf Bottom}) By including lensing power spectra, the sensitivity to $\sigma_0$ is now stronger for the nonthermal case and cold ICs than the thermalized case. \label{fig:S4bounds} }
\end{figure}

\FloatBarrier
\clearpage

\section{Effect of DM-baryon scattering on matter power spectrum}

In the main text, we neglected DM-baryon drag effects when obtaining limits on freeze-in from the matter power spectrum. Fig.~\ref{Pk_with_drag} shows the effect on the matter power spectrum when both the phase space and DM-baryon drag are included (solid lines), for the case where the DM sector is thermalized. DM-baryon drag reduces the matter power spectrum slightly at larger scales (smaller $k$), since the effect is largest at late times around the epoch of recombination (see Fig.~\ref{dragrate}). Meanwhile, the exponential cutoff in the matter power spectrum is primarily set by smaller scales, corresponding to modes which entered the horizon when the DM was semi-relativistic. It can be seen from Fig.~\ref{Pk_with_drag} that the half-mode suppression scale is thus largely unaffected by DM-baryon drag, justifying our approach in the main text. Nevertheless, as noted before, using the half-mode suppression gives only approximate bounds and a robust analysis of freeze-in requires accounting for the full transfer function.

\begin{figure}[t]
    \centering
    \includegraphics[width=0.6\textwidth]{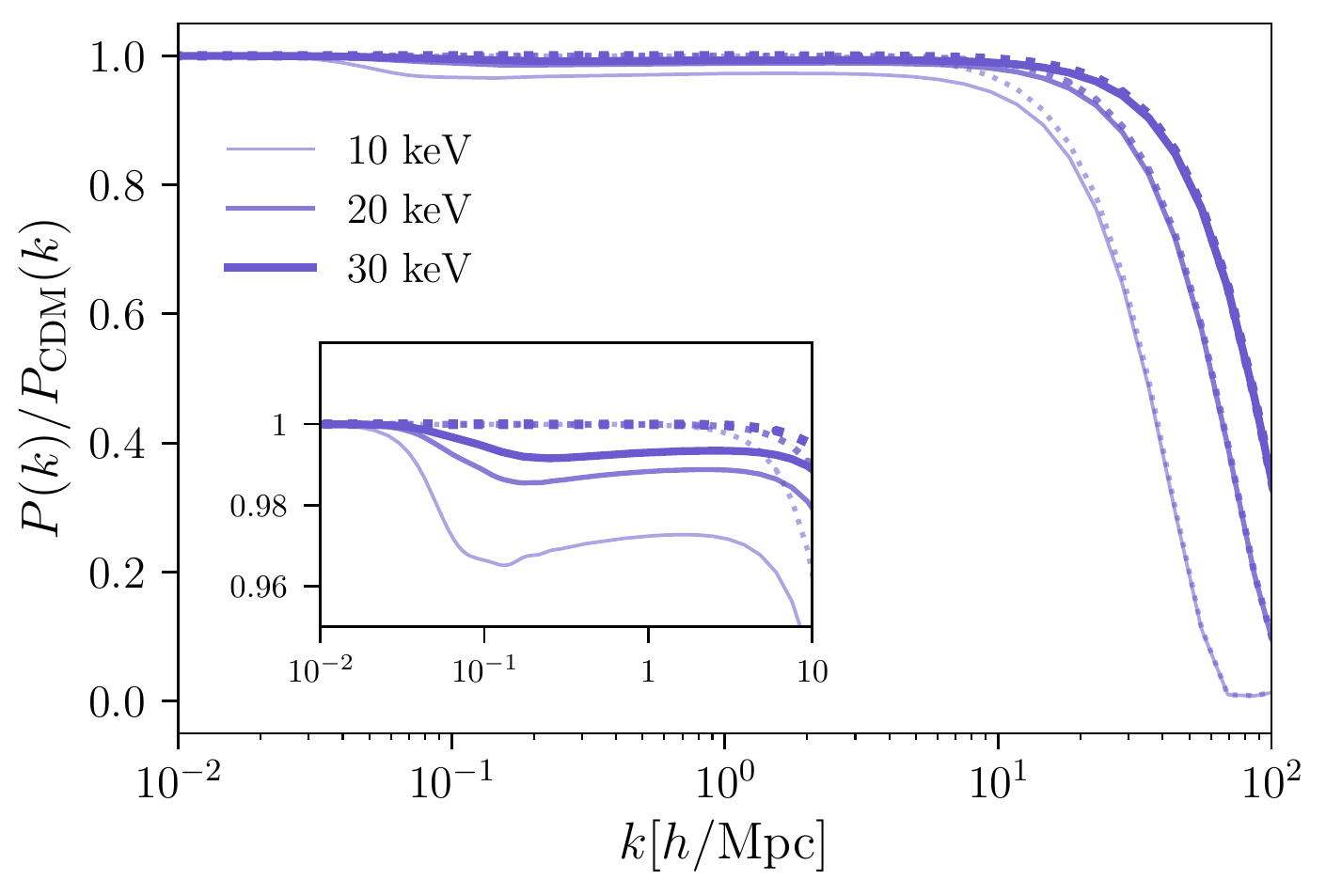}
    \caption{Comparison of the transfer fraction with DM-baryon drag (solid) or turning off the drag (dotted). The drag does not affect the half-mode suppression scale, but it does cause an additional slight suppression to the transfer function at large scales (inset). Here we have assumed freeze-in DM which has thermalized at early times. The cross sections are set by the couplings needed to obtain the relic density.}
    \label{Pk_with_drag}
\end{figure}

\section{Dark Photon Parameter Space for Freeze-in}
In Fig.~\ref{darkphoton}, we show the relevant parameter space if freeze-in is mediated by a dark photon with mass $m_{A'}\ll m_\chi$ that is kinetically mixed with the SM photon. We impose a DM-mass dependent constraint on DM self-interactions using Eq.~(4) of Ref.~\cite{Dvorkin:2019zdi} and requiring that $\sigma_T/m_\chi<1\,$cm$^2/$g at the velocity scale relevant for merging clusters. This translates to an upper bound on $g_\chi$ and therefore a lower bound on $\kappa$ for DM made by freeze-in at a given mass. For DM as heavy as 1~MeV, the dark photon can be as heavy as $\sim 10^{-3}$~eV, whereas for DM in the mass range we constrain, $m_\chi\sim 10$~keV, the dark photon must be significantly lighter. 
\begin{figure}[t]
    \centering
    \includegraphics[width=0.7\textwidth]{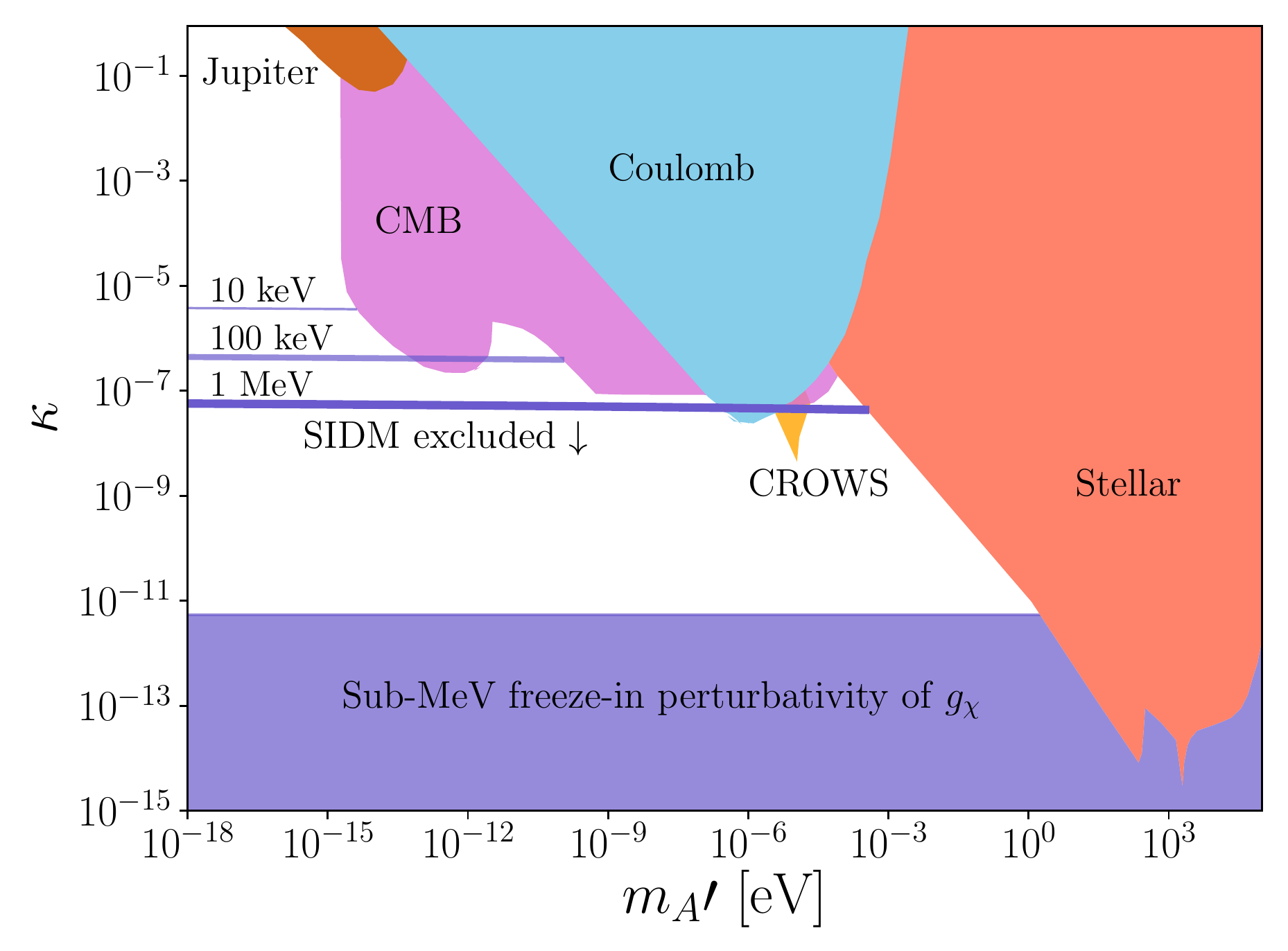}
\caption{Parameter space of interest if sub-MeV freeze-in is mediated by a kinetically mixed dark photon that is significantly lighter than the DM (making freeze-in an infrared-dominated process). The excluded region below $\kappa \sim 10^{-11}$ is specific to freeze-in; this is the value of saturation where $\kappa = eQ$ at $m_\chi=1$~MeV, which would require a non-perturbative coupling $g_\chi>1$. Tighter constraints that are specific to freeze-in come from requiring that $g_\chi$ not be too large so as to violate bounds on self-interacting DM (SIDM), for instance from the bullet cluster~\cite{Tulin:2017ara}. Regions above the horizontal lines, which correspond to self-interaction bounds for 10~keV, 100~keV, and 1~MeV DM masses, are allowed for the freeze-in mechanism to make all the DM. Further model-independent constraints on the parameter space, shown as shaded regions, come from Ref.~\cite{Jaeckel:2013ija} and have been updated with newer bounds from the CMB~\cite{Caputo:2020bdy} as well as bounds from the CERN resonant weakly interacting sub-eV particle search (CROWS)~\cite{Betz:2013dza}. Note also that black hole superradiance constrains the region between $\sim 10^{-14} - 10^{-11}$~eV and between$ \sim 10^{-19} - 10^{-17}$~eV in the limit of small couplings (i.e. where gravitational interactions are dominant in the dynamics of the black hole cloud)~\cite{Baryakhtar:2017ngi}; the exact value of $\kappa$ where gravitational interactions become dominant has not yet been established, which is why we do not include the superradiance bound on the figure.}
    \label{darkphoton}
\end{figure}
\end{document}